\documentclass[10pt,letterpaper]{article}
\usepackage[top=0.85in,left=2.75in,footskip=0.75in]{geometry}

\usepackage{amsmath,amssymb, amsthm}
\theoremstyle{definition}
\newtheorem{definition}{Definition}[section]
\usepackage{changepage}

\usepackage[utf8x]{inputenc}
\usepackage{textcomp,marvosym}
\usepackage{cite}
\usepackage{nameref,hyperref}
\usepackage[right]{lineno}

\usepackage{microtype}
\DisableLigatures[f]{encoding = *, family = * }

\usepackage[dvipsnames]{xcolor}
\usepackage{array}
\usepackage{ulem}
\newcolumntype{+}{!{\vrule width 2pt}}

\newlength\savedwidth

\raggedright
\usepackage[english]{babel}
\usepackage[aboveskip=1pt,labelfont=bf,labelsep=period,justification=raggedright,singlelinecheck=off,figurename=Fig]{caption}

\makeatletter
\renewcommand{\@biblabel}[1]{\quad#1.}
\makeatother

\usepackage{lastpage,fancyhdr,graphicx}
\usepackage{epstopdf}
\usepackage{nameref}
\fancyheadoffset[L]{2.25in}
\fancyfootoffset[L]{2.25in}
\usepackage{subcaption}
\usepackage{algorithm}
\usepackage{algpseudocode}
\usepackage{enumerate}
\graphicspath{{img_2/}}

\begin{document}
\vspace*{0.2in}

\begin{flushleft}
{\Large
\textbf\newline{Determining clinically relevant features in cytometry data using persistent homology} 
}
\newline
\\
Soham Mukherjee\textsuperscript{1,\Yinyang},
Darren Wethington\textsuperscript{2,4,\Yinyang},
Tamal K. Dey\textsuperscript{1},
Jayajit Das\textsuperscript{2,3,4,5*}
\\
\bigskip
\textbf{1} Computer Science Department, Purdue University, West Lafayette, Indiana, USA
\textbf{2} Biomedical Science Graduate Program, Wexner College of Medicine, The Ohio State University, Columbus, Ohio, USA
\textbf{3} Biophysics Program, The Ohio State University, Columbus, Ohio, USA
\textbf{4} Battelle Center for Mathematical Medicine, Abigail Wexner Research Institute, Nationwide Children's Hospital, Columbus, Ohio, USA
\textbf{5} Departments of Pediatrics, Biomedical Informatics, and, Pelotonia Institute of Immuno-Oncology, Wexner College of Medicine, The Ohio State University, Columbus, Ohio, USA \\
\bigskip

\Yinyang These authors contributed equally to this work.\\
* jayajit@gmail.com

\end{flushleft}

\section*{Abstract}
Cytometry experiments yield high-dimensional point cloud data that is difficult to interpret manually. Boolean gating techniques coupled with comparisons of relative abundances of cellular subsets is the current standard for cytometry data analysis. However, this approach is unable to capture more subtle topological features hidden in data, especially if those features are further masked by data transforms or significant batch effects or donor-to-donor variations in clinical data. We present that persistent homology, a mathematical structure that summarizes the topological features, can distinguish different sources of data, such as from groups of healthy donors or patients, effectively. Analysis of publicly available cytometry data describing non-na{\"i}ve CD8+ T cells in COVID-19 patients and healthy controls shows that systematic structural differences exist between single cell protein expressions in COVID-19 patients and healthy controls. We identify proteins of interest by a decision-tree based classifier, sample points randomly and compute persistence diagrams from these sampled points. The resulting persistence diagrams identify regions in cytometry datasets of varying density and identify protruded structures such as `elbows'. We compute Wasserstein distances between these persistence diagrams for random pairs of healthy controls and COVID-19 patients and find that systematic structural differences exist between COVID-19 patients and healthy controls in the expression data for T-bet, Eomes, and Ki-67.  Further analysis shows that expression of T-bet and Eomes are significantly downregulated in COVID-19 patient non-na{\"i}ve CD8+ T cells compared to healthy controls. This counter-intuitive finding may indicate that canonical effector CD8+ T cells are less prevalent in COVID-19 patients than healthy controls. This method is applicable to any cytometry dataset for discovering novel insights through \emph{topological data analysis} which may be difficult to ascertain otherwise with a standard gating strategy or existing bioinformatic tools.

\section*{Author summary}
Identifying differences between cytometry data seen as a point cloud can be complicated by random variations in data collection and data sources. We apply \emph{persistent homology} used in \emph{topological data analysis} to describe the shape and structure of the data representing immune cells in healthy donors and COVID-19 patients. By looking at how the shape and structure differ between healthy donors and COVID-19 patients, we are able to definitively conclude how these groups differ despite random variations in the data. Furthermore, these results are novel in their ability to capture shape and structure of cytometry data, something not described by other analyses.


\section{Introduction}

Cytometry data contain information about the abundance of proteins in single cells and are widely used to determine mechanisms and biomarkers that underlie infectious diseases and cancer. Recent advances in flow and mass cytometry techniques enable measurement of abundances of over 40 proteins in a single cell~\cite{spitzer2016mass,simoni2018mass}. Thus, in the space spanned by protein abundance values measured in cytometry experiments, a cytometry dataset is represented by a point cloud composed of thousands of points where each point corresponds to a single cell. Abundances of proteins or chemically modified forms (e.g., phosphorylated forms) of proteins in single immune cells change due to infection of the host by pathogens (e.g., a virus) or due to the presence of tumors which usually result in changes in the `shape' of point cloud data measured in cytometry experiments~\cite{mathew2020deep,strauss2015human,wargo2016monitoring}.
Cytometry data analysis techniques commonly rely on Boolean gating and calculation of relative proportions of resulting populations as a method to compare datasets across control/healthy and experimental/diseased conditions. In recent years, state-of-the-art analyses based on sophisticated machine learning algorithms capable of mitigating batch effects, ad hoc gating assumptions, and donor-donor variability have been developed~\cite{azad, del_barrio_optimalflow:_2020}. However, these methods are not designed to quantitatively characterize shape features (e.g., connected clusters, cycles) in high dimensional cytometry datasets that can contain valuable information regarding unique co-dependencies of specific proteins in diseased individuals compared to healthy subjects.

Topological Data Analysis (TDA) aims to capture the underlying shape of a given dataset by describing its topological properties. Unlike geometry, topological features (e.g., the hole in a doughnut) are invariant under continuous deformation such as rotation, bending, twisting but not tearing and gluing. One of the tools by which TDA describes topological features latent in data is persistent homology~\cite{edelsbrunner2010computational,zomorodian2012topological}.
For example, for a point cloud data, persistent homology captures the birth and death of topological features (e.g., `holes') in a dataset after building a scaffold called a simplicial complex out of the input points. This exercise provides details regarding topological features that `persist' over a range of scale and thus contain information regarding the shape topology at different length scales (see \ref{fig:s1}~Fig for details).  Persistent homology has been applied to characterize shapes and shape-function relationships in a wide variety of biological systems including skin pattern formation in zebra fish~\cite{zebrafish}, protein structure, and pattern of neuronal firing in mouse hippocampus~\cite{komendantov2019quantitative}. TDA has additionally previously been applied to identify immune parameters associated with transplant complications for patients undergoing allogenic stem cell transplant using populations of immune cell types assayed via mass cytometry~\cite{lakshmikanth2017mass}. 
However, this work did not use persistent homology or expression levels of proteins in their analysis, leaving the shape of  cytometry data uncharacterized. Another work focuses on the use of TDA as a data reduction method for single-cell RNA sequencing data~\cite{rizvi2017single}, but again do not attempt to characterize how topologies derived from point clouds  differ among  disparate data sources such as healthy and diseased individuals.

The challenges of directly applying current persistence methodologies to cytometry data to characterize distinguishing features between healthy and diseased states are the following: 1. Features that separate healthy from diseased state can pertain to the change in density of points in a region in point cloud data - therefore, the information of local density should be incorporated in persistent homology methods, in particular in the filtration step that brings in sequentially the simplices connecting the points. In commonly used Rips filtration~\cite{buchet2016efficient} the density of points is not included. 2. There can be shape changes giving a different length scale in the point cloud data, such as formation of an elbow, in a diseased condition. 3. There can be systematic differences between healthy and diseased states across batch effects and donor-donor variations. Topological features should capture these global differences being oblivious to the local variations caused
by measurement noise.

We address the above challenges by developing an appropriate filtration function to compute persistence and applying the method to characterize distinguishing features of non-na{\"i}ve CD8+ T cells between healthy and SARS-CoV-2 infected patients.  

\section{Results}
\subsection{Persistence framework for  SARS-CoV-2 infection}
\label{sec:results}
Topological signatures given by persistence are stable, global, scale invariant and show resilience to local perturbations~\cite{cohen2007stability}. It is this property of persistent homology that motivates us to use TDA in distinguishing clinically relevant features in flow cytometry data in COVID-19 patients.

\begin{figure}[!hbt]
    \centering
    \includegraphics[width=0.9\linewidth]{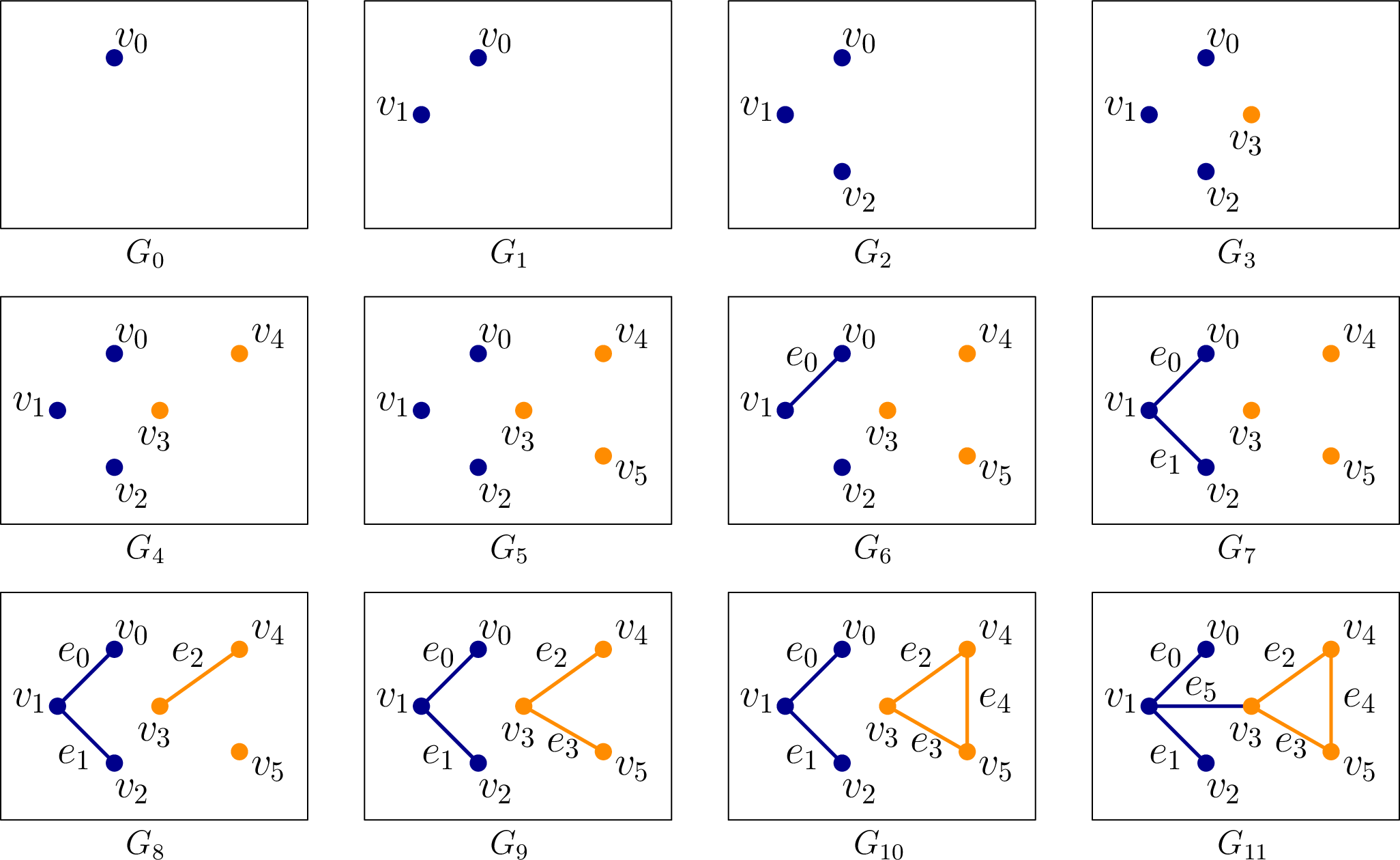}
    \caption{{\bf An example of \emph{filtration} for a graph}. The nested sequence of graphs
    $G_0\subset G_1\subset \ldots G_{11}$ forms a filtration of the final graph $G_{11}$. Each vertex $v_i$ creates a new component in the nested sequence, and edges $e_0, e_1, e_2, e_5$ merge two components whereas $e_4$ creates a cycle (yellow).}
    \label{fig:graph_filtration}
\end{figure}

\paragraph{Persistent Homology:}Persistent homology builds on 
an algebraic structure called homology groups graded by its dimension $i$ and  denoted by $\sf{H}_i$ . Intuitively, they describe the shape of the data by `connectivity' at
different levels. For example, $\sf{H}_0$ describes the number of connected components, $\sf{H}_1$ describes the number of holes, and, $\sf {H}_2$ describes the number of enclosed voids apparently present in the shape that the dataset represents. Three and higher dimensional homology groups capture analogous higher ($\geq 3$) dimensional features. A point cloud data (henceforth abbreviated as PCD) itself does not have much of a `connected structure'. So, a scaffold called a \emph{simplicial complex} is built on top of it. This simplicial complex, in general, is made out of simplices of various dimensions such as vertices, edges, triangles, tetrahedra, and other higher dimensional analogues. Given a growing sequence of such complexes called \emph{filtrations}, a persistence algorithm tracks information regarding the homology groups across this sequence. In our case, these complexes can be restricted only to vertices and edges. 
With the restriction that both vertices of an edge appear before
the edge, we get a nested sequence of graphs 
$$ G_0 \subset G_1 \subset G_2 \subset \ldots G_n$$ as the filtration. Fig~\ref{fig:graph_filtration} shows such a \emph{filtration}.

\begin{figure}[!htb]
    \centering
    \includegraphics[width=0.95\linewidth]{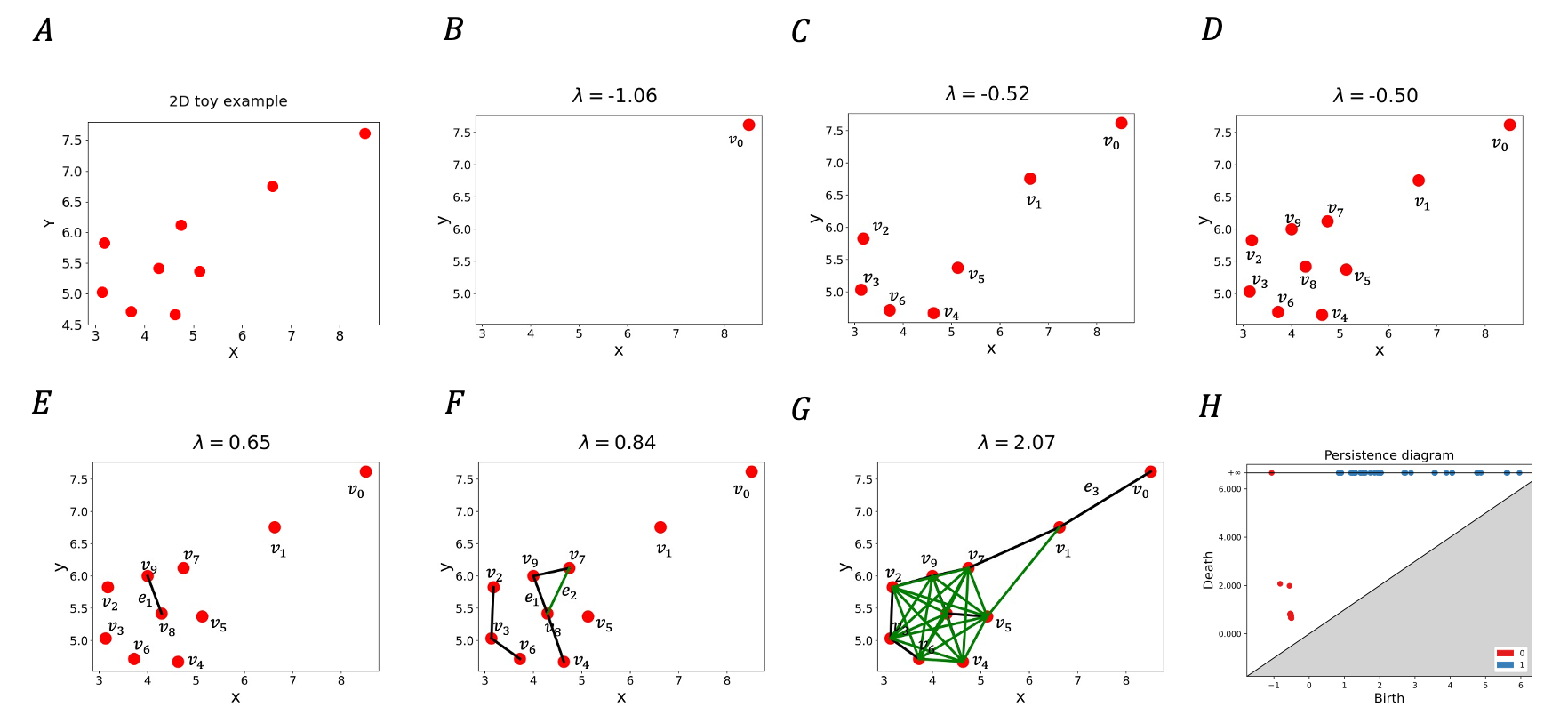}
    \caption[Illustration of persistence for a 2D point cloud data (PCD)]%
    {{\bf Illustration of persistence for a 2D point cloud data (PCD)}.
    {\bf (A), (H)} shows a $2$D PCD example and its computed persistence diagram. {\bf(B)-(G)} shows important changes in topological feature as $\lambda$ increases from $-\infty$ to $\infty$. {\bf(B)} At $\lambda = -1.06$ an isolated point, $v_0$ appears first. Note that each isolated vertex creates a new component. {\bf(C)} At $\lambda = -0.52$ points in the denser region appears in the filtration, introducing more components. The indices of the vertices denote the order in which they appear in the filtration. {\bf(D)} At $\lambda = -0.50$, all vertices appear in the filtration. Note that, the way we have chosen the filtration function $f$, vertices appear before the edges since $f_v(v)$ is always negative. {\bf(E)} At $\lambda = 0.65$, the first edge $e_1$ appears merging two components. By persistence algorithm~\cite{edelsbrunner2010computational}, we pair the edge $e_1$ with $v_9$, since $v_9$ appears later in the filtration. Corresponding to this, we get a persistence pair $(f_v(v_9), f_e(e_1)) = (-0.50, 0.65)$. {\bf(F)} At $\lambda = 0.84$, the green edge $e_2$ appears and creates a cycle. Since there is no 2-simplex(triangle) present, the cycle is never destroyed. In the persistence diagram we have this pair as $(f_e(e_3), \infty) = (0.84, \infty)$. {\bf (G)} At $\lambda = 2.07$, the long edge $e_3$ appears joining $v_0$ and $v_1$, yielding a persistence pair $(-1.06, 2.07)$.}
    \label{fig:intuition}
\end{figure}

\paragraph{Persistence Diagram:} Appearance (`birth') and disappearance (`death') of topological features, that is, cycles whose classes constitute the homology groups, can be captured by persistence algorithms~\cite{edelsbrunner2010computational, carlsson2005persistence}. These `birth' and `death' events are represented as points in the so-called \emph{persistence diagram}. If a topological feature is born at filtration step $b$ and dies at step $d$, we represent this by \emph{persistence pair} $(b, d)$ with persistence $d-b$. The pair $(b,d)$ becomes a point in the persistence diagram with the `birth' as x-axis and `death' as y-axis. This 2D plot summarizes topological features latent in the data. In the example-filtration of Fig~\ref{fig:graph_filtration}, a new component gets `born' when a vertex $v_i$ appears in the filtration for the first time. When an edge is introduced, one of the two things can happen--either two components are joined, or a cycle is created. In the first case, a `death' happens for $0$-th homology group $\sf{H}_0$, and in the second case, a `birth' happens for the
$1$-st homology group $\sf{H}_1$. For example, when $e_0$ comes in the filtration ($G_6$), it merges two components created by $v_0$ and $v_1$. By convention, we choose to kill the component that got created later in the filtration and thus we let the component created by $v_1$ die. We obtain a persistence pairing $(1, 6)$ since edge $e_0$ at filtration step $6$ kills the component created by $v_1$ at step $1$. Similarly, we obtain pairs $(2, 7), (4 ,8), (5, 9), \text{ and } (3, 11)$. These points, tracking the `birth' and `death' of components, produce
the persistence diagram for the $0$-th homology group ${\sf H}_0$ and hence
we refer to it as the ${\sf H}_0$-persistence diagram.
Note that the edge $e_4$ creates a cycle (yellow) that never dies. In such cases, \textit{i.e.} when a topological feature never dies, we pair it with $\infty$. For the edge $e_4$, we obtain a persistence pair $(10, \infty)$. But, this
feature concerns the $1$-st homology group ${\sf H}_1$ and thus it becomes
a point in the persistence diagram for ${\sf H}_1$ which we refer to as
${\sf H}_1$-persistence diagram.
One way to leverage the above framework for studying a function 
is to assign function values to vertices and edges and construct a filtration by ordering them according to these assigned values. For such cases the persistence pairs take the form $(b,d)$ where $b$ is the value at which a feature is born and $d$ is the value at which it dies. The function values
that induce the filtration (Fig~\ref{fig:intuition}) are chosen to capture two features of the input PCD--(i) the density variations, and (ii) the \emph{anisotropy} of the features, that is, how elongated it is
in a certain direction, henceforth termed as \emph{length scale} `of the feature' or collectively `of the data'. In particular, length scales refer to the prominence of protrusions such as `elbows' in COVID-19 data.

Below we briefly describe how we adapt the above persistence framework for analyzing point cloud data (PCD) representing CD8+ T cells in SARS-CoV-2 infection. Details regarding the approach are provided in Section~\ref{sec:methods} and \ref{sec:alg} Appendix and \ref{fig:s1}~Fig.

\paragraph{Computing persistent homology for cytometry datasets:} 
Our datasets consist of cytometry data for non-na{\"i}ve CD8+ T cells.
Given protein expressions (real values) for $d$ proteins in such a single cell, we can represent it as a $d$-dimensional point in $\mathbb{R}^d$. Considering a population of single cells, we get a point cloud (PCD) in $\mathbb{R}^d$. Now, we study the shape of this PCD using the persistence framework that we describe above. We compute persistence diagrams for the PCDs generated with protein expressions from different individuals and compare them.
It turns out that, for computational purposes, we need a limit on the dimension $d$ for PCD which means we need to choose carefully the proteins that differentiate effectively the subjects of our 
interest, namely the healthy individuals, COVID-19 patients, and recovered patients. We typically choose 3 (sometimes 2) protein expressions to generate the PCD and call it a PCD in the  \emph{P1, P2, P3 space} if it is generated by proteins P1, P2, and
P3 respectively.

Flow cytometry data for non-na{\"i}ve CD8+ T cells in Mathew et al.~\cite{mathew2020deep} show generation of CD8+ T cells with larger abundances of the proteins CD38 and HLA-DR (CD38+HLA-DR+ cells) for some COVID-19 patients, forming an `elbow' in the two dimensional PCD with CD38 and HLA-DR protein expressions (see \ref{fig:s2}~Fig). Moreover, there is an increase in the local density of the points (or single CD8+ T cells) in the `elbow' region. 
This suggests that, to study the PCD generated by the protein expressions by persistence framework, we need to choose a filtration that is able to capture such geometric shapes and variations in the local density. 

We briefly describe our choice of filtration by considering the example of a point cloud $P\subset\mathbb{R}^2$ shown in Fig~\ref{fig:intuition}. Mathematical and computational details regarding the filtration are provided in the Section \ref{sec:methods}.  
We build a filtration according to
assigned values to the vertices and edges of
a graph connecting the input points. For a vertex $p$ which is a point in the input PCD $P$, we denote this value $f_v(p)$ (given by Eq~\ref{eq:fil1} in Section~\ref{sec:methods}). Similarly, we denote the assigned value to an edge $e$ as $f_e(e)$ (given by Eq~\ref{eq:fil2} in Section~\ref{sec:methods}); see Fig~\ref{fig:intuition}.
The values satisfy the conditions that $f_v(p) < 0$ and $f_e(e)\geq 0$; implications of this specific choice will become clear in the next paragraph.   
It is noteworthy to mention that $f_v(p)$ is the negative of distance-to-measure originally defined in~\cite{chazal} and later used in~\cite{buchet2017declutter} for the PCD case and captures the density distribution of points,  whereas $f_e(e)$ captures the inter-point distances between the points in the given point cloud.

The persistence algorithm processes each vertex and edge in the order of their appearance in the filtration. We execute it using a threshold value $\lambda$ from $-\infty$ to $\infty$ and generate the persistence diagram accordingly. Intuitively, as $\lambda$ is increased from $-\infty$ to $\infty$, vertices $p$ for which $f_v(p)\leq \lambda$  and edges for which $f_e(e)\leq \lambda$  appear in the filtration for a particular value of $\lambda$ (see Fig~\ref{fig:intuition}). Since  $f_v(p) < 0$ and $f_e(e)\geq 0$, all the vertices first appear as $\lambda$ is increased from $-\infty$ to $0$, and then edges start appearing as $\lambda$ becomes positive. The birth-death events for $\sf{H}_0$ and $\sf{H}_1$ constituting the persistence diagram (Fig~\ref{fig:intuition}H) contain information about the density and length scales present in the point-cloud. For example, the points showing birth and death events for the $\sf{H}_0$-persistence diagram are more densely organized for the single cell protein expression data from the healthy donor than the SARS-CoV-2 infected patient in the HLA-DR - CD38 plane shown in \ref{fig:s3}~Fig. The denser organization of the birth-death events in the persistence diagram indicates a more homogeneous distribution of CD38 and HLA-DR proteins in the CD8+ T cells in healthy donors compared to that in infected patients. Most of the CD8+ T cells in healthy controls have low amounts of CD38 and HLA-DR abundances and few contain larger values of these proteins, indicating a greater degree of homogeneity. The birth-death events for $\sf{H}_1$ in the persistence diagram (\ref{fig:s3}~Fig) in general contain information about the length scales of cyclic structures in the point cloud. It also can capture protrusions like `elbows' that we have in COVID-19 data. Our filtration allows only birth (and not death) of 1-cycles and therefore, a $\lambda$ value corresponding to the birth of a 1-cycle captures the length scale of the newly born cycle and hence an `elbow'. Our analysis of the PCDs in Fig~\ref{fig:s3}D and \ref{fig:s3}H indeed shows that $\lambda$ values for the birth of cycles for the COVID-19 patient is much larger compared to that for the healthy individual indicating the presence of larger length scales in the PCD which is consistent with the presence of an `elbow' shape in the PCD for the patient. 

\subsection{Application of persistence to healthy and patient data}
Our aim is to find out systematic differences in topological features extracted from cytometry data for healthy individuals and COVID-19 patients. Ideally one would like to compute persistence diagrams for all 25 proteins that were measured in single CD8+ T cells, however, this task encounters two major problems. First, as we mentioned before taking the full 25 dimensional PCD introduces the \textit{curse of dimensionality}~\cite{bellman1966dynamic} making it computationally infeasible to produce the persistence diagrams. The second one is more subtle. In order to measure how the density of data differs from a healthy to infected person in a quantitative way, we need to ensure that the number of points in each PCD, to be analyzed by persistent homology, is the same. Cytometry data usually contain different numbers of single cells in datasets obtained from different donors or replicates. To address the \textit{curse of dimensionality} we use a classifier (XGBoost~\cite{chen2016xgboost}) that distinguishes single CD8+ T cells in healthy donors from those in COVID-19 patients and we choose the top $r$ (taken to be 3) features (proteins) that are deemed important by the classifier while classifying the data points (cells). This reduces the dimension of the data from 25 to a much smaller value denoted $r$. 

\begin{figure}[!t]
    \centering
    \includegraphics[width=\linewidth]{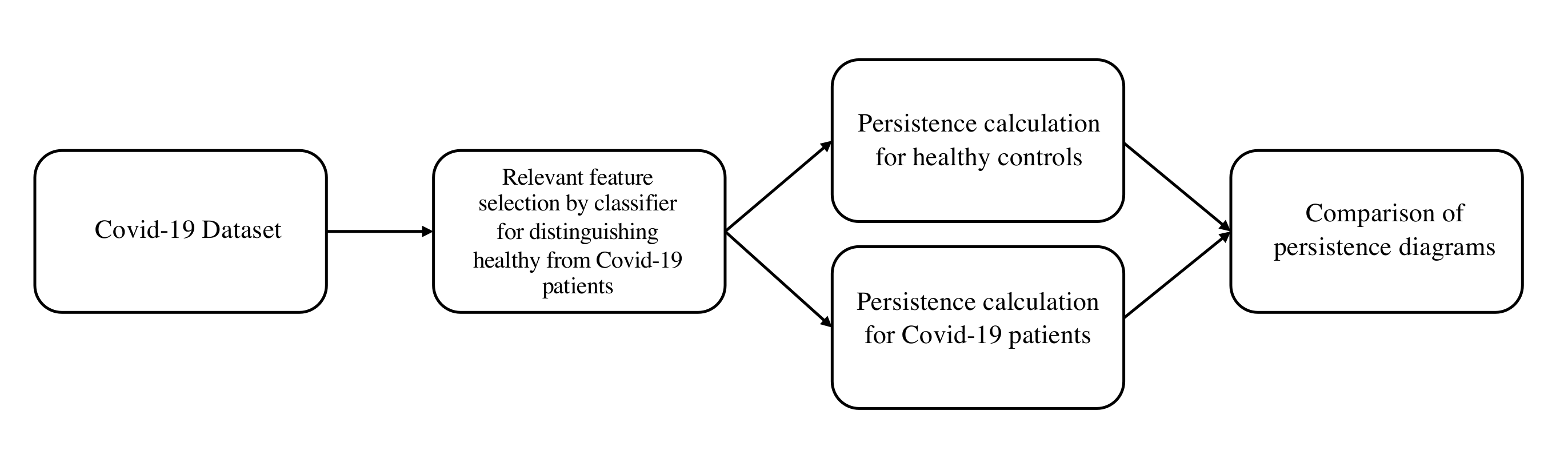}
    \caption{{\bf Flowchart of computation pipeline}. The pipeline includes three main stages, namely, (i) relevant feature selection, (ii) persistence computation, and (iii) comparison of persistence diagrams.}
    \label{fig:method} 
\end{figure}

To address the second issue, we perform uniform random sampling on every $r$-dimensional dataset and take equal number of samples from it. We then use the filtration defined in Eq~\ref{eq:fil1} and Eq~\ref{eq:fil2} to construct persistence diagrams for each dataset. To quantify the structural differences in the datasets as captured by the corresponding persistence diagrams, we compute the Wasserstein distance~\cite{kerber2017geometry} between persistence diagrams from randomly selected pairs of either two healthy donors (H$\times$H) or a healthy donor and an infected patient (H$\times$P) and compute distributions of the Wasserstein distances for a large number of (H$\times$H) and (H$\times$P) pairs. The comparison of these distributions via Kolmogorov-Smirnov (KS) tests provides information regarding the systematic differences in shape features in the CD8+ T cell cytometry data across healthy individuals and COVID-19 patients. The computational pipeline is summarized in (Fig~\ref{fig:method}).
Below we describe results from the application of our computational pipeline to the CD8+ T cell cytometry data in Mathew et al.~\cite{mathew2020deep} 

\begin{figure}[!hbt]
    \centering
    \includegraphics[width=0.9\linewidth]{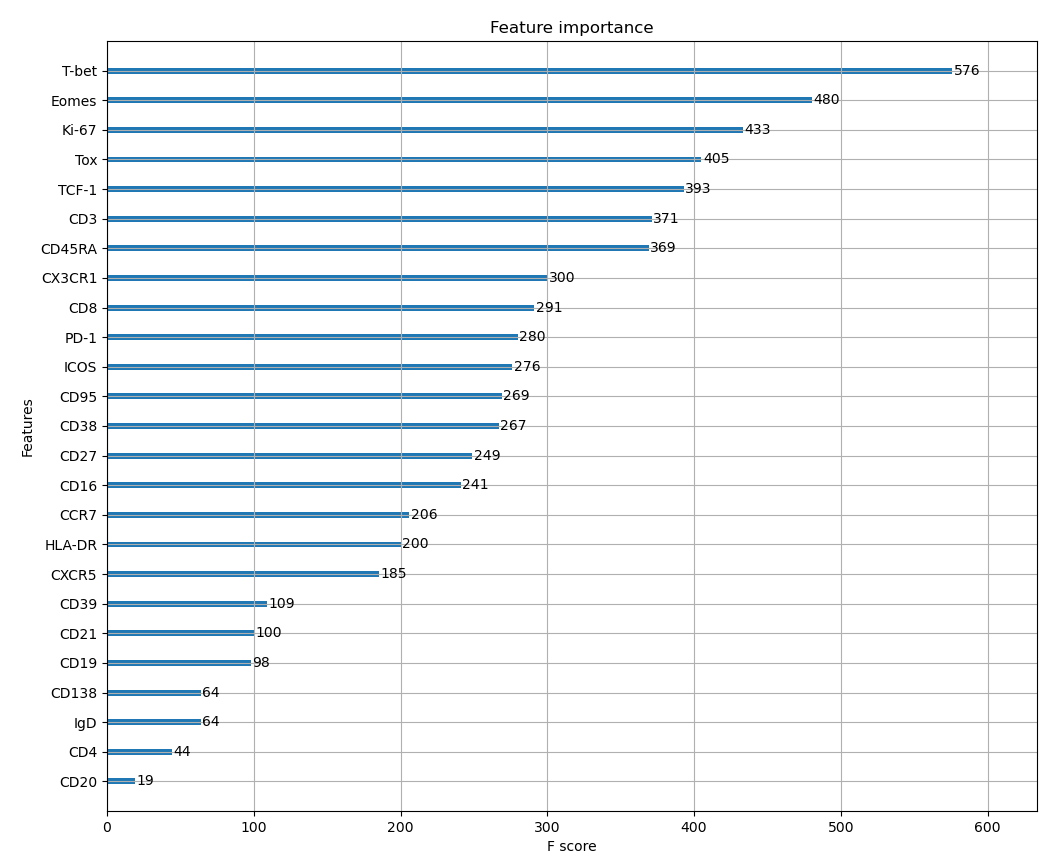}
    \caption{{\bf Rank ordering of proteins using a decision tree based classifier}. Shows rank ordering of proteins by descending values of feature importance generated by the classifier XBoost.}
    \label{fig:feature-imp}
\end{figure}

\paragraph{A few protein expressions in CD8+ T cells separate healthy donors from COVID-19 patients:} We use XGBoost~\cite{chen2016xgboost}, a decision tree based classifier, to rank order proteins for their ability to distinguish CD8+ T cell point cloud data between healthy individuals and COVID-19 patients. The average accuracy of the classifier is about $92\% $. The classifier returns a feature score for each protein that characterizes its importance relative to other proteins in distinguishing cells from healthy individuals and COVID-19 patients. Intuitively, feature score is an indicator of the importance of a particular feature while classifying the data. By ranking the proteins by their feature scores, we can reduce our further analysis to only a subset of the most important proteins. Our analysis (Fig~\ref{fig:feature-imp}) shows that the top three most important proteins to the XGBoost classifier are proteins T-bet, Eomes, and Ki-67. T-bet induces gene expressions leading to an increase in cytotoxic functions of CD8+ T cells. CD8+ T cells with increased cytotoxic functions are known as `effector' CD8+ T cells and these cells show higher T-bet abundances. Conversely, Eomes induces gene expressions that contribute towards increased life span and re-activation potential of CD8+ T cells to specific antigens~\cite{takemoto2006cutting}. These long-lived T cells are known as `memory' T cells which show increased expressions of Eomes. Memory T cells provide key protection against re-exposure to the same infection. Ki-67 is a marker for actively proliferating cells~\cite{scholzen2000ki}. Mathew et al.~\cite{mathew2020deep} identified Ki-67 as one marker that is upregulated (increased) in some COVID-19 patients. These three proteins are most likely to distinguish CD8+ T cells in healthy donors from those in patients. Further details regarding the application of the classifier are provided in the Materials and Methods section (Section~\ref{sec:methods}). 
 
\begin{figure}[!hbt]
    \centering
    \includegraphics[width=0.9\linewidth]{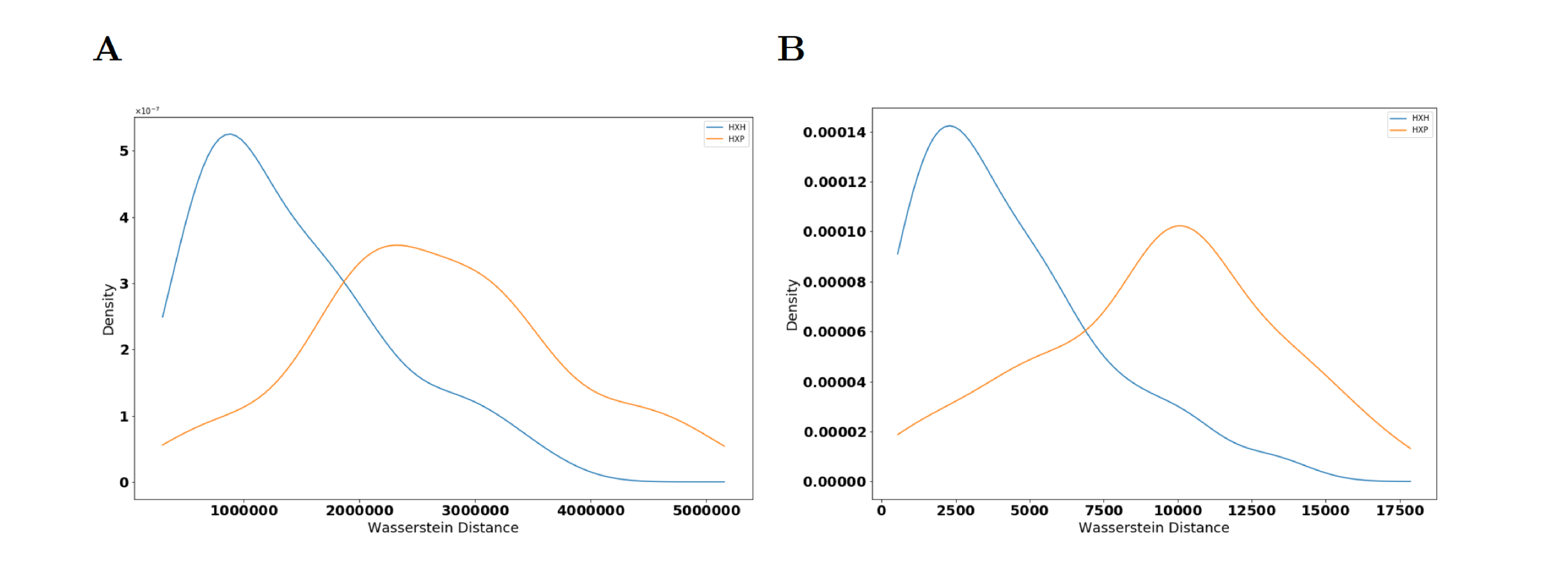}
    \caption{{\bf Distributions of Wasserstein distances between persistence diagrams}. {\bf(A)} Shows distributions of Wasserstein distance between ${\sf H}_0$-persistence diagrams for H$\times$H (blue line) and H$\times$P (orange line) pairs ($p=8.77\times10^{-15}$, $QFD=0.173$). {\bf (B)} Shows distributions of Wasserstein distance between ${\sf H}_1$-persistence diagrams for H$\times$H (blue line) and H$\times$P (orange line) for the same pairs in (A) ($p=3.04\times10^{-14}$, $QFD=0.219$). Persistence diagrams are calculated from point clouds in the T-bet, Eomes, and Ki-67 axes. p-values are calculated from a 2-sided KS test.}
    \label{fig:wasserstein}
\end{figure}

\paragraph{Persistence diagrams distinguish structural features in CD8+ T cell data occurring in healthy individuals and COVID-19 patients across batch effects and donor-donor variations:} We select the proteins T-bet, Eomes, and Ki-67 as relevant markers and compute the persistence diagrams of the PCD given by them for each individual belonging to groups of healthy donors, COVID-19 patients, and recovered patients. The persistence diagrams vary from individual to individual in each group and between groups which could arise due to batch effects in samples and/or donor-to-donor variations. To determine if there are systematic differences in persistence diagrams for individuals across the three groups (healthy, patient, and recovered), we compute Wasserstein distance between persistence diagrams for 3 categories of pairings: 1) two healthy donors (H$\times$H), 2) one healthy donor and one patient (H$\times$P), and 3) one healthy donor and one recovered individual (H$\times$R). We compute distances for $100$ randomly chosen pairs of individuals for each category of pairings. Wasserstein distances of the persistence diagrams for $0$-th and $1$-st homology groups $\sf{H}_0$ and $\sf{H}_1$ respectively are higher when comparing H$\times$P pairs than when comparing H$\times$H pairs (Fig~\ref{fig:wasserstein}). A 2-sided KS test showed that the Wasserstein distances for H$\times$P and H$\times$H belong to different probability distribution functions ($p\ll 0.01$); see Fig~\ref{fig:wasserstein} and also the description of this test in Section~\ref{sec:methods}. This indicates that systematic geometric differences in the flow cytometry PCD with T-bet, Eomes, and Ki-67 between individuals with and without COVID-19 are not attributable to batch effects or donor-to-donor variations alone. Increasing the number of randomly chosen pairs to $200$ did not change this conclusion as Figs~\ref{fig:wasserstein} and \ref{fig:s4} illustrate. The difference between H$\times$H and H$\times$R distributions of distances in the T-bet, Eomes, and Ki-67 space are less prominent (\ref{fig:s5}~Fig). We further test if such systematic differences are present for proteins that are at the bottom of the list in Fig~\ref{fig:feature-imp} and find that the distributions of Wasserstein distances for corresponding persistence diagrams overlap between the H$\times$H and H$\times$P pairs (\ref{fig:s6}~Fig). This suggests that systematic differences in the geometry of the PCD occur only for specific sets of proteins. Details regarding computation of persistence diagrams and Wasserstein distances are given in Section~\ref{sec:methods}.

\begin{figure}[!htb]
    \centering
    \includegraphics[width=0.9\linewidth]{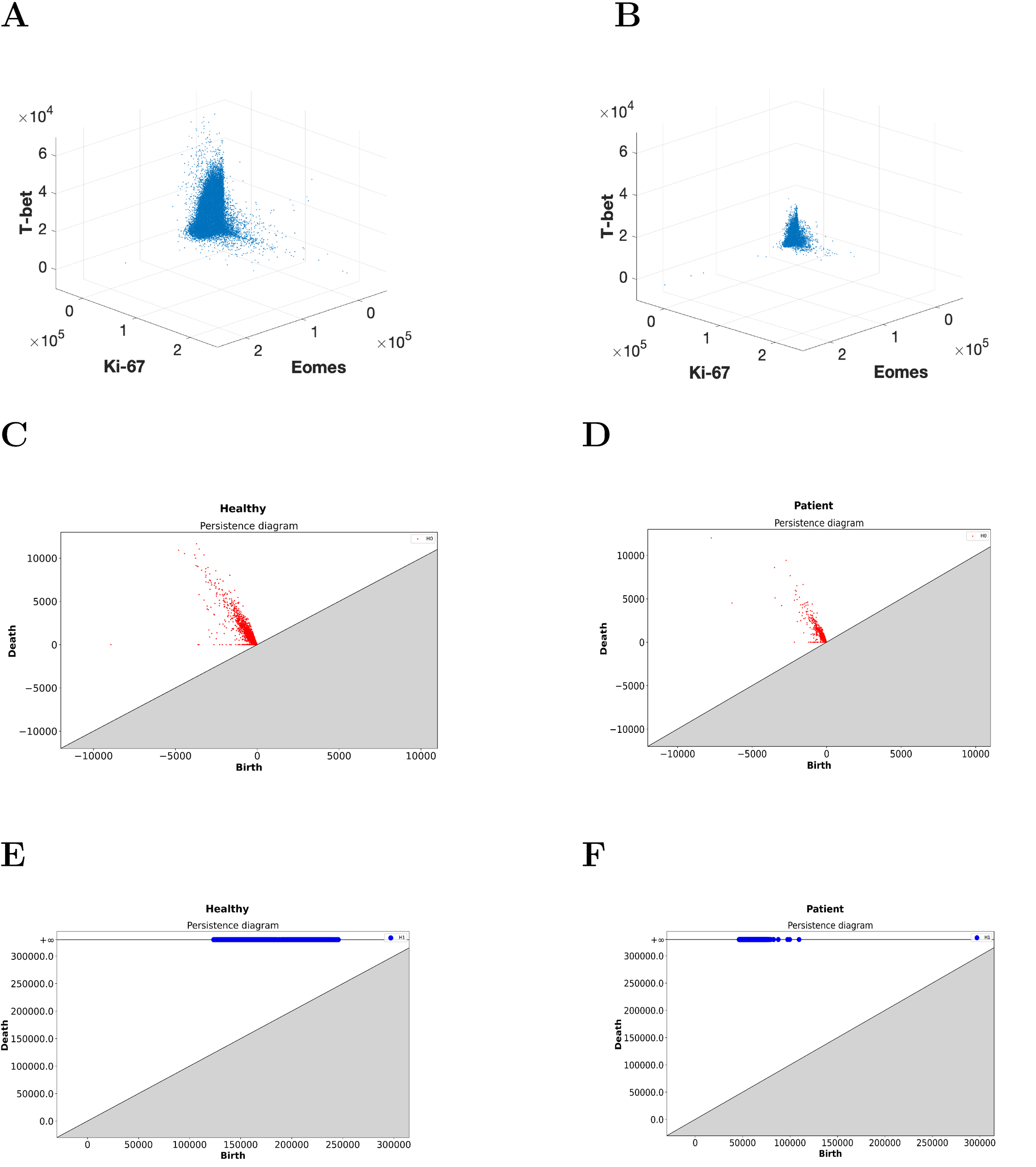}
    \caption{{\bf Differences in shape features in the 3D point cloud for CD8+ T cells in a H$\times$P pair}.  CD8+ T cell point cloud for proteins Eomes, Ki-67, and T-bet for \textbf{(A)} a healthy control and \textbf{(B)} a COVID-19 patient. \textbf{(C)} Shows ${\sf H}_0$-persistence diagram for the healthy control in (A). \textbf{(D)} Shows the ${\sf H}_0$-persistence diagram for the COVID-19 patient in (B). \textbf{(E)} ${\sf H}_1$-persistence diagram for the healthy control in (A). \textbf{(F)} ${\sf H}_1$-persistence diagram for the COVID-19 patient in (B).}
    \label{fig:data_and_pers}
\end{figure}

Next, we select a comparison pair that generates a large Wasserstein distance between ${\sf H}_1$-persistence diagrams to further investigate what structural differences exist between the datasets. We choose one pair of a healthy control and patient that generated a Wasserstein distance of $4.0\times10^6$ units in their ${\sf H}_0$-persistence diagrams and $1.1\times10^4$ units in ${\sf H}_1$-persistence diagrams. These two individual PCDs and their resulting persistence diagrams are shown in Fig~\ref{fig:data_and_pers}.

A readily apparent difference between the resulting persistence diagrams is given by the lower birth times in ${\sf H}_1$ of the COVID-19 patient compared to the healthy control (Fig~\ref{fig:data_and_pers}E and \ref{fig:data_and_pers}F). This result indicates that the length scale of the data is smaller in the COVID-19 patient, which can be visually confirmed in the scatter plots of the data (Fig~\ref{fig:data_and_pers}A and \ref{fig:data_and_pers}B). Specifically, the single cell abundances of T-bet and Eomes in CD8+ T cells are clustered significantly tighter around the origin for the COVID-19 patients than for the healthy controls.   Similar manual inspection of other H$\times$P pairs that generate large Wasserstein distances between their persistence diagrams confirms that this trend is not limited to this pair alone. 

Additionally, the points in the ${\sf H}_0$-persistence diagram are spread out more widely for the healthy control than the COVID-19 patient (Fig~\ref{fig:data_and_pers}C and \ref{fig:data_and_pers}D). A wider distribution of births and deaths in the $0$-th homology ${\sf H}_0$ implies that there are regions of disparate densities. This suggests that the densities in the protein expressions of T-bet and Eomes are more homogeneous in the PCD in the COVID-19 patient than in the healthy control.

The structural change in the PCD for CD8+ T cells in the T-bet/Eomes plane that occurs during COVID-19 infection implies that T-bet and Eomes expression should be downregulated (decreased) in non-na{\"i}ve CD8+ T cells. This result is consistent with analysis of clusters of CD8+ T cells by Mathew et al.~\cite{mathew2020deep} via a software package FlowSOM~\cite{van2015flowsom} that shows that clusters high in T-bet and/or Eomes are downregulated in COVID-19 patients.

The relevance of the above proteins in distinguishing healthy controls from patients is further demonstrated by the statistically significant differences (p-values $\ll 10^{-8}$) in the mean T-bet, Eomes, and Ki-67 abundances in the CD8+ T cells between the groups (\ref{fig:s10}~Fig). However, the distributions of the mean abundances for the above proteins also showed regions of overlap between healthy and patient populations (\ref{fig:s10}~Fig) indicating existence of $H\times P$ pairs with much smaller differences in the mean values between them than the population averaged mean values of these proteins. Our method specifically identifies differences in topological features in the shape of the PCD between a $H\times P$ pair which can be present despite small differences in the mean values of specific proteins (e.g., T-bet). We further investigated this point by analyzing correlations between the Wasserstein distance between the persistence diagrams of $H\times P$ pairs with the difference in the mean protein abundances (Fig~\ref{fig:s11}A and \ref{fig:s11}B) which showed moderate correlations ($\leq 0.5$). However, there are several instances in which the Wasserstein distance captures differences in the shape of the PCD via persistent homology even when the difference in mean protein abundance (e.g., mean T-bet abundance) is small (Fig~\ref{fig:s11}C-\ref{fig:s11}F).  This is due to changes in the shape of the point-cloud that are not easily captured by summary statistics such as the mean. Therefore, our analysis shows that healthy and COVID-19 pairs can be better separated by our persistent homology analysis than by summary statistics measures in such cases. The downregulation of T-bet and Eomes in response to viral infections is not well documented, as CD8+ T cells commonly differentiate into phenotypes with high T-bet, high Eomes, or both in response to infections~\cite{knox2014characterization,takemoto2006cutting}.

\begin{figure}[!hbt]
    \centering
    \includegraphics[width=\linewidth]{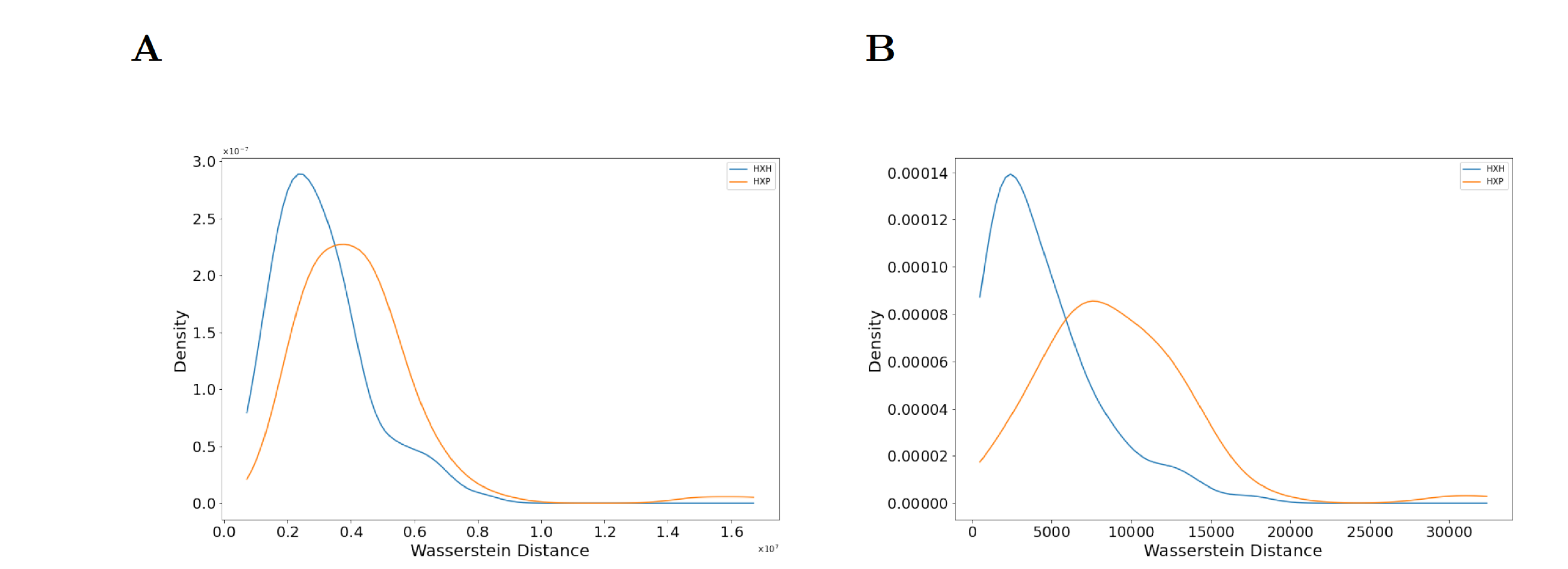}
    \caption{{\bf Distributions of Wasserstein distances between persistence diagrams for B cells}. {\bf(A)} Distributions of Wasserstein distance between ${\sf H}_0$-persistence diagrams for H$\times$H (blue line) and H$\times$P (orange line) pairs (p=$6.28 \times 10^{-5}$, QFD=$0.0300$). {\bf (B)} Distributions of Wasserstein distance between ${\sf H}_1$-persistence diagrams for H$\times$H (blue line) and H$\times$P (orange line) for the same pairs in (A) (p=$1.07 \times 10^{-12}$, QFD=$0.0946$). Persistence diagrams are calculated from point clouds in the CXCR5, PD-1, and TCF-1 axes. p-values are calculated from a 2-sided KS test.}
    \label{fig:b_cells}
\end{figure}

We next explore the application of our approach to other datasets. We apply our method to the single cell cytometry dataset in Mathew et al.~\cite{mathew2020deep} for B cells obtained from healthy donors and COVID-19 patients. The B cells are major orchestrators of the humoral component of adaptive immunity against infections. We compute persistence diagrams for the proteins CXCR5, PD-1, and TCF-1, identified by XGBoost as the three most important proteins for classifying healthy donors and patients. A chemokine receptor, CXCR5, is responsible for B cell trafficking and is found to be downregulated in B cells in COVID-19 patients~\cite{mathew2020deep}. Both PD-1, a checkpoint inhibitory receptor~\cite{thibult2013pd}, and TCF-1, a transcription factor important for T cell differentiation and effector functions~\cite{fimmu} are increased in B cells in infected individuals (\ref{fig:s12}~Fig). Immunosuppressive effects of high PD-1 expression in B cells have been reported earlier~\cite{thibult2013pd}. We find that Wasserstein distances of the persistence diagrams for both homology groups (${\sf H}_0$ and ${\sf H}_1$) are significantly different (Fig~ \ref{fig:b_cells}) between the healthy donors and the patient population B cells. This demonstrates that our approach is able to distinguish healthy individuals from patient populations using PCDs of other immune cells. Furthermore, we find that the margin of separation, quantified by the QF-distance (QFD), in these Wasserstein distances is smaller with B cells than the CD8+ T cells. This implies that the structure of PCDs for CD8+ T cells differ more between healthy controls and patients than that for the B cells. These findings may point to previously uncharacterized impact of PD-1 and TCF-1 on B cell function or phenotype in SARS-CoV-2 infection. 

Distributions of mean PD-1 expression on B cells in COVID-19 patients is not largely different than that of healthy controls (\ref{fig:s12}B~Fig). Therefore, we further analyzed how the difference between mean PD-1 values and the differences in PCDs quantified by the Wasserstein distances are related (Fig~\ref{fig:s13}A-\ref{fig:s13}D). We found that unlike T-bet for CD8+ T cells, mean PD-1 expression differences in B cells are not tightly correlated with Wasserstein distances in healthy control-patient pairs (Fig~\ref{fig:s13}A-\ref{fig:s13}B). We visualized the PCD in the space of TCF-1, PD-1, and CXCR5 (Fig \ref{fig:s13}E-\ref{fig:s13}F) to gain further insights regarding the differences in the shapes of the PCD in healthy control-patient pairs with similar mean PD-1 expressions. The differences in the shape of the PCDs for these pairs can be largely attributed to the higher expressions of TCF-1 in healthy controls (Fig~\ref{fig:s13}E-\ref{fig:s13}F).

\paragraph{Comparison with Existing Methods:}
To determine how our selection of filtration compares with an existing method, we compare how our results might change if we use Rips filtration. In Rips filtration, simplices appear when all of their edges appear in the filtration. The edges appear in non-decreasing order of their lengths. We use the entire dataset as the PCD and generate persistence diagrams subsequently, using Rips filtration~\cite{edelsbrunner2010computational,buchet2016efficient,dey_wang_2019} rather than the filtration
we use in our approach. Note that in the standard Rips filtration, all vertices appear at the same instant whereas in our case the vertices are ordered by Eq~\ref{eq:fil1}. We then compute Wasserstein distances as done previously. We find that Rips filtration is also able to distinguish persistence diagrams of healthy controls and patients, but the margin of separation is much lower, as evidenced by a higher p-value and lower QFD than our choice of filtration offers (\ref{fig:s14}~Fig). This indicates that our method, which is designed to identify protrusions such as ``elbows" in the data, characterizes greater differences in CD8+ T cell protein expression structures than existing methods such as Rips filtration.

We then compare our TDA approach with an existing algorithm FlowSOM~\cite{van2015flowsom}, widely used for visualizing, clustering, and analyzing PCDs from cytometry experiments. FlowSOM uses a self-organizing map algorithm for generating single cell subsets with unique marker protein expressions. FlowSOM is capable of clustering similar cells together and offers a robust way to determine which cellular subsets are differentially expressed between data sources~\cite{quintelier2021analyzing}. We run a FlowSOM analysis and clustering on the CD8+ T cell data and determine that 6 of the 15 clusters are differentially expressed between healthy controls and COVID-19 patients (\ref{fig:s15}~Fig).

\begin{figure}[!hbt]
    \centering
    \includegraphics[width=\linewidth]{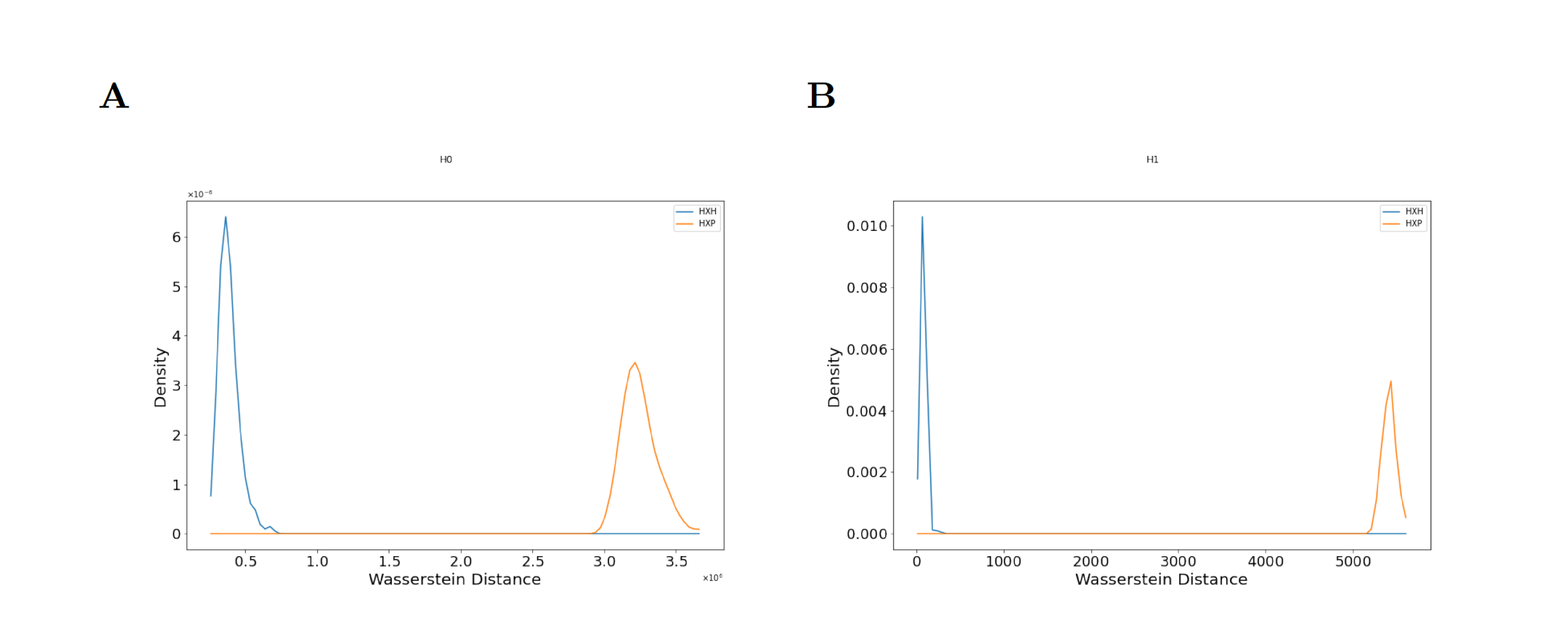}
    \caption{{\bf Distributions of Wasserstein distances between persistence diagrams from a FlowSOM cluster (Cluster ${\#}$1) that is not differentially expressed between healthy controls and COVID-19 patients}. {\bf(A)} Distributions of Wasserstein distance between ${\sf H}_0$-persistence diagrams for H$\times$H (blue line) and H$\times$P (orange line) pairs  (p=$2.21 \times 10^{-59}$, QFD=$1.638$) computed for the PCD for Eomes, Ki-67, and T-bet for CD8+ T cells. {\bf (B)} Distributions of Wasserstein distance between ${\sf H}_1$-persistence diagrams for H$\times$H (blue line) and H$\times$P (orange line) for the same pairs in (A) (p=$2.21 \times 10^{-59}$, QFD=$1.903$).}
    \label{fig:flowsom_cluster1}
\end{figure}

We select one FlowSOM cluster which is differentially expressed (Cluster \#3) and one that is not differentially expressed (Cluster \#1) between healthy donors and patients for downstream analysis. Visual inspection of these clusters in the 3-dimensional space of T-bet, Eomes, and Ki-67 shows that PCD structure may be different between healthy controls and patients in Cluster \#1 (\ref{fig:s15}C~Fig). This is because proteins can co-vary in different ways in healthy controls and patients, affecting topological features hidden in the PCD. We then perform our persistent homology analysis to determine if the structure of PCDs for proteins Eomes, Ki-67, and T-bet in subsets of single cells associated with these FlowSOM clusters differ between healthy controls and patients. We find distributions of Wasserstein distances between the persistence diagrams obtained for the above FlowSOM clusters are significantly different (Figs~\ref{fig:flowsom_cluster1} and \ref{fig:s16}).

Performing FlowSOM on B cells reveals 12 of the 15 clusters are differentially expressed between healthy controls and COVID-19 patients (\ref{fig:s17}A-\ref{fig:s17}B Fig). The three clusters (Cluster \#1, \#2, and \#14) that are not differentially expressed are all high in PD-1. Visualization of the PCDs for Cluster \#2 and Cluster \#4 shows that regions high in TCF-1 \ref{fig:s17}C-\ref{fig:s17}D Fig) distinguish the PCDs for the pairs. Thus, our method is able to identify topological features hidden in the PCD that separate healthy controls from COVID-19 patients in FlowSOM clusters, including clusters which are not differentially expressed between these groups.

\section{Discussions and conclusions} 

We develop a persistent homology based approach to determine topological features hidden in point cloud data representing single cell protein abundances measured in cytometry data. In particular, we characterize the number of connected components or ${\sf H}_0$, and the number of holes or ${\sf H}_1$ in our persistence calculations, and show that our approach is able to determine systematic shape differences in the cytometry data for CD8+ T cells obtained from healthy individuals and COVID-19 patients. Therefore, the approach is able to successfully determine systematic shape differences that exist in the presence of batch effect noise and donor-donor variations in the cytometry data. Furthermore, our approach does not use data transformations (e.g., arc-sinh transformation) or any ad-hoc subtype gating to determine these systematic differences, thus we expect persistent homology based approaches will be especially useful in identifying high-dimensional structural trends hidden in cytometry data.

We determine structural changes in T-bet and Eomes abundances in single CD8+ T cells in COVID-19 patients that can be summarized as downregulation. This result is non-intuitive as previous findings show that T-bet and Eomes protein abundances are highest in effector CD8+ T cells, which are induced in response to acute infections, suggesting T-bet and Eomes expressions should be upregulated~\cite{knox2014characterization,takemoto2006cutting}. The clinical implications of this result are unclear. Mathew et al.~\cite{mathew2020deep} describe a immunophenotype in which Eomes+, T-bet+, CD8+ T cells are more abundant in COVID-19 patients who respond poorly to Remdesevir and NSAIDs, have high levels of IL-6, and have fewer eosinophils. Our analysis identifies that this immunophenotype (i.e.,Eomes+, T-bet+, CD8+ T cells) is systematically less prevalent in COVID-19 patients than in healthy controls. The ability of our approach to identify shape features in single immune cell PCD without any `supervision' (e.g., specific gating) of the cytometry data shows that it can potentially determine more complicated immunologically relevant shape features. Furthermore, our approach inferred finer
geometric structures from PCDs in B cells for proteins CRCX5, PD-1, and TCF-1 which helped distinguish COVID-19 patients from healthy individuals. These proteins are associated with cell migration, immunosuppression, and effector functions in lymphocytes and can potentially provide further insights into B cell response in COVID-19.

We compare our approach with an existing algorithm FlowSOM which is widely used for analyzing and visualizing multidimensional cytometry data. Our comparison reveals PCDs for subtypes of CD8+ T cells in FlowSOM clusters that do not differentiate healthy controls and COVID-19 patients contain topological features separating the above groups. Therefore, detecting topological features hidden in the PCDs can provide important biological insights regarding response of the lymphocytes in COVID-19.

Our approach integrates cellular comparisons with dataset comparisons. First, the classifier pools all data and determines which proteins are significant in discriminating whether cells come from healthy controls or COVID-19 patients. In this way, the classifier identifies a way to compare cellular phenotypes across experimental groups. Next, the computation of Wasserstein distances for persistence diagrams compares individuals against each other, integrating cellular phenotypes with donor information (e.g., healthy and COVID-19 patients). Thus, this approach allows us to automatically identify individuals that are associated with distinguishing structural features in the point cloud data.

Currently, the limitations are mostly due to the curse of dimensionality that increases the computational complexity. Since we are computing pairwise distances between datapoints to obtain the persistence diagram (\ref{sec:alg}~Appendix), computation time increases as the dimension of data increases. Computational cluster resources that we use currently complete all computations in about 20 minutes. This is comparable to other data science applications using large datasets, but this approach can be a barrier to those without access or experience with computational clusters. Additionally, it is unclear how additional dimensions impact the statistical properties of the data and interpretability of the results. To expand into many (i.e. 25) dimensions, computational interpretation and validation tools are necessary.

\section{Materials and methods} 
\label{sec:methods}
\paragraph{Relevant feature selection by the XGBoost classifier:}

Let $D = \big \{ c_1, c_2, \ldots, c_m \big \}$ be the collection of $m$ cytometry datasets.
 Each dataset, $c_i$, can be viewed as a $\mathcal{M}_{n\times p}$ matrix where $n$ is the number of datapoints (cells) and $p$ is the number of proteins with which each $c_i$ is generated. We denote the collection of cytometry datasets of healthy individuals as $C_{\mathcal{H}} \subset D$ and similarly the set of individuals infected with SARS-CoV-2 as $C_{\mathcal{P}} \subset D$. We proceed to label the data in the following manner: If $c_i \in C_{\mathcal{H}}$ then we assign the label $+1$ to each of the $n$ datapoints, similarly we assign $-1$ if  $c_i \in C_{\mathcal{P}}$. Essentially, we now have a binary classification problem where our labeled dataset is $D^{\prime} = \bigcup D = c_1 \cup c_2 \cup \ldots \cup c_j$, with labels defined as above. We solve this binary classification problem with XGBoost~\cite{chen2016xgboost}, a gradient boosted decision tree based classifier, and as a byproduct we get feature scores that correspond directly to each feature's importance in the classification. The higher the score for a protein, the more important it is for the classifier's decision. After our classifier orders the proteins  by their scores, we take first $r$ proteins to construct the point-cloud on which persistence diagrams are computed. We set $r=3$ for all our analysis reported here. We used data from $56$ healthy individuals and $108$ COVID-19 patients for our feature selection.

The XGBoost classifier was implemented using the open-source python XGBoost package~\cite{chen2016xgboost}. The model was then trained and validated with $K$-fold cross-validation, with $K=10$. The average accuracy of the classifier was \textbf{$92.14 \pm 0.04 \%$}. The protein scores are shown in Fig~\ref{fig:feature-imp}. 

\paragraph{Random subsampling of the datapoints:} Each PCD can be thought as a set of indexed points. These indices were first shuffled randomly and then $20,000$ indices and hence respective points were sampled uniformly from this shuffled set. The samples drawn from each PCD were further analyzed using persistent homology. We discarded datasets that had less than $20,000$ data points. Among $55$ healthy individuals only $1$ had less than $20,000$ data points. For the patient data, the number of such datasets was $34$.
\paragraph{Details of persistent homology computation:} 
 As mentioned before (Section~\ref{sec:results}), computation of persistence diagrams needs a \emph{filtration.} 
We set the filtration induced by the function $f=\{f_v, f_e\}$ where $f_v(p)$ computes an `average' Euclidean distance between the vertex $p$ and its $k$ neighbors according to Eq~\eqref{eq:fil1} and $f_e(e)$ computes the length of the edge $e$ according to Eq~\eqref{eq:fil2}.
\begin{align}
    f_v(p) &= -\frac{1}{k} \sqrt{\sum_{i}^{k} \Vert p - q_i \Vert^2}\text{  },p \in P, \text{ and } q_i \in k\text{-Nearest Neighbors of } p. \label{eq:fil1} \, .
\end{align}
 The term $\Vert p - q_i\Vert$ in the above equation is the Euclidean distance between the vertices $p$ and $q_i$. The function value $f_e(e)$ for an edge $e=(p,q)$ is given by the Euclidean distance between $p$ and $q$. For the experiments, the number of nearest neighbors is fixed to $k = 40$.
\begin{align}
f_e(e) &= \Vert p - q \Vert\text{ , } \forall p, q \in P \text{ and } p \neq q \label{eq:fil2}
\end{align}

We begin by sampling every cytometry PCD $c_i$ and take $n(= 20,000)$ samples. We do this to make $c_i$ uniform \textit{w.r.t.} number of data points (single CD8+ T cells). We compute a complete weighted graph $G(V, E)$ with vertices in the sampled data. This complete graph $G$ is a key-step that enables us to compute the \emph{persistence diagram}, $Dgm(c_i)$ of the dataset $c_i$, \textit{w.r.t.} the filtration function $f$. We show the algorithm (Algorithm~B in \ref{sec:alg} Appendix) that executes this step in detail in the supplementary material. Notice that the graph $G$ is weighted in the sense that each vertex $v \in V$ and edge $e \in E$ carries a weight of $f_v(v)$ and $f_e(e)$ respectively. Observe that $f:V \cup E \rightarrow \mathbb{R}$ constitutes a valid filtration of $G$.

 We compute persistence diagrams for each $c_i\in D$ according to Algorithm~C in \ref{sec:alg} Appendix. The next step involves comparing the persistence diagrams. We do this by computing the Wasserstein distance between persistence diagrams and plotting their distributions. We take two persistence diagrams of randomly selected healthy individuals and compute the Wasserstein distance between them with the help of Gudhi~\cite{gudhi:urm,kerber2017geometry} and scikit-learn Python library~\cite{sklearn_api}. Similarly, we compute Wasserstein distance between persistence diagrams of a healthy and an infected individual (both are randomly drawn from the collection). We plot the resulting distances. We do this for 108 pairs to obtain two distributions. Note that, results described in Section~\ref{sec:results} still holds for $200$ pairs (\ref{fig:s4}~Fig). Intuitively, a large Wasserstein distance between two persistence diagrams implies the datasets on which they were constructed are structurally very different while a small distance implies they are structurally similar.

\paragraph{Statistical testing of difference in Wasserstein distance distributions:} 2-sided Kolmogorov–Smirnov (KS) tests were performed on Wasserstein distances for pairs of individuals to determine if they arise from the same or different probability distribution functions~\cite{mit}. MATLAB's subroutine \textit{kstest2} was used to determine p-values, where the null hypothesis is that the Wasserstein distances from H$\times$H comparisons and the experimental condition (either H$\times$P or H$\times$R) arise from the same non-parametric distribution, and the alternative hypothesis is that they come from different distributions. A $p$-value of $\leq 0.05$ (or $\geq 0.05$) indicates the support for the alternate hypothesis (i.e., data occur from different distributions) is statistically significant (or not significant).

\paragraph{Computing quadratic form (QF) distance:}
To measure the dissimilarity between a pair of Wasserstein distance distributions, quadratic form (QF) distance was computed as proposed by Bernas et al. in~\cite{bernas2008quadratic} (originally introduced in~\cite{Hafner}). The QF distance was calculated using the formula
\begin{equation}
    D^2(\mathbf{h}, \mathbf{f}) =(\mathbf{h}-\mathbf{f})^T \mathbf{A}_{j}^{i} (\mathbf{h}-\mathbf{f}) =  \sum_{i=1}^{n}\sum_{j=1}^{n}a_{ij}(h_i - f_i) (h_j - f_j)
\end{equation}
where $\mathbf{f}$ and $\mathbf{h}$ are two vectors that list counts corresponding to two histogram bin counts. The quantities $\mathbf{f}$ and $\mathbf{h}$ can be normalised so that $\sum_{i}f_i = \sum_{i} h_i = 1$ when indexed by $i$. In our case, $\mathbf{A}_j^{i} = [a_{ij}]$ and defined as $a_{ij} = 1 - \sqrt{\frac{(i-j)^2}{d_{max}}}$ with $d_{max}$ being maximum distance between bins.

\paragraph{FlowSOM Analysis:}
FlowSOM was performed on the entire non-na{\"i}ve CD8+ dataset, and separately, the non-na{\"i}ve B cell dataset. Data was scaled with the transform asinh(x/150) before analysis. See Code Availability for FlowSOM source code. Comparisons of relative cluster abundances between healthy controls and COVID-19 patients were performed with a Wilcoxon rank sum test. Subsequent persistent homology computation was performed on the selected clusters by sampling $20,000$ cells from either the aggregated healthy control data or aggregated COVID-19 patient data. This sampling was repeated to form ``synthetic" individual healthy control or COVID-19 patient data. We cannot sample data from each individual, as was done in the prior computations, because many individuals display too few cells in the selected clusters to reliably sample $20,000$ cells. In the CD8+ T cell analysis, clusters \#1 and \#3 were chosen for persistent homology calculations because they contain the most cells, and thus are most likely to have cells in each sample and be unaffected by random sampling.

\paragraph{Flow cytometry data for healthy individuals and COVID-19 patients:} The data come from Mathew et al., 2020 \cite{mathew2020deep} and was retrieved from Cytobank. Mathew et al. performed high-dimensional flow cytometry experiments using peripheral blood obtained from 125 patients admitted to the hospital with COVID-19, 36 donors that recovered from documented SARS-CoV-2 infection, and 60 healthy controls. Our analysis focuses on the deposited data available at \url{https://premium.cytobank.org/cytobank/experiments/308357} for non-na{\"i}ve CD8+ T cells collected at the time of admission (and not any later blood draws, such as at 7 days after admission). Please note that a Cytobank account is currently required for data access. We removed forward- and side-scatter variables and other non-protein measurements, resulting in 25 proteins included in our analysis.

\paragraph{Available code:}
\label{sec:code} Our current code is available at \url{https://github.com/soham0209/TopoCytometry} and will be updated for ease-of-use and performance enhancements.

\paragraph{Acknowledgments:} This work is supported by the NIH awards R01-AI 143740 and R01-AI 146581 to JD, and by the Research Institute at the Nationwide Children's Hospital, and NSF awards CCF 1839252 and 2049010 to TD. The authors would like to thank John Wherry and members of his lab for depositing their data and helping us to understand it.  

\nolinenumbers

\begin{thebibliography}{10}

\bibitem{spitzer2016mass}
Spitzer M, Nolan G.
\newblock Mass Cytometry: Single Cells, Many Features.
\newblock Cell. 2016;165(4):780--791.
\newblock doi:{https://doi.org/10.1016/j.cell.2016.04.019}.

\bibitem{simoni2018mass}
Simoni Y, Chng MHY, Li S, Fehlings M, Newell EW.
\newblock Mass cytometry: a powerful tool for dissecting the immune landscape.
\newblock Current Opinion in Immunology. 2018;51:187--196.
\newblock doi:{https://doi.org/10.1016/j.coi.2018.03.023}.

\bibitem{mathew2020deep}
Mathew D, Giles JR, Baxter AE, Oldridge DA, Greenplate AR, Wu JE, et~al.
\newblock Deep immune profiling of COVID-19 patients reveals distinct
  immunotypes with therapeutic implications.
\newblock Science. 2020;369(6508).
\newblock doi:{10.1126/science.abc8511}.

\bibitem{strauss2015human}
Strauss-Albee DM, Fukuyama J, Liang EC, Yao Y, Jarrell JA, Drake AL, et~al.
\newblock Human NK cell repertoire diversity reflects immune experience and
  correlates with viral susceptibility.
\newblock Science Translational Medicine. 2015;7(297):297ra115--297ra115.
\newblock doi:{10.1126/scitranslmed.aac5722}.

\bibitem{wargo2016monitoring}
Wargo JA, Reddy SM, Reuben A, Sharma P.
\newblock Monitoring immune responses in the tumor microenvironment.
\newblock Current Opinion in Immunology. 2016;41:23--31.
\newblock doi:{https://doi.org/10.1016/j.coi.2016.05.006}.

\bibitem{azad}
Azad A, Rajwa B, Pothen A.
\newblock Immunophenotype Discovery, Hierarchical Organization, and
  Template-Based Classification of Flow Cytometry Samples.
\newblock Frontiers in Oncology. 2016;6:188.
\newblock doi:{10.3389/fonc.2016.00188}.

\bibitem{del_barrio_optimalflow:_2020}
del Barrio E, Inouzhe H, Loubes JM, Matrán C, Mayo-Íscar A.
\newblock {optimalFlow}: optimal transport approach to flow cytometry gating
  and population matching.
\newblock BMC Bioinformatics. 2020;21(1):479.
\newblock doi:{10.1186/s12859-020-03795-w}.

\bibitem{edelsbrunner2010computational}
Edelsbrunner H, Harer J.
\newblock Computational topology: an introduction.
\newblock American Mathematical Soc.; 2010.

\bibitem{zomorodian2012topological}
Zomorodian A.
\newblock Topological data analysis.
\newblock Advances in applied and computational topology. 2012;70:1--39.

\bibitem{zebrafish}
McGuirl MR, Volkening A, Sandstede B.
\newblock Topological data analysis of zebrafish patterns.
\newblock Proceedings of the National Academy of Sciences.
  2020;117(10):5113--5124.
\newblock doi:{10.1073/pnas.1917763117}.

\bibitem{komendantov2019quantitative}
Komendantov AO, Venkadesh S, Rees CL, Wheeler DW, Hamilton DJ, Ascoli GA.
\newblock Quantitative firing pattern phenotyping of hippocampal neuron types.
\newblock Scientific Reports. 2019;9(1):1--17.
\newblock doi:{10.1038/s41598-019-52611-w}.

\bibitem{lakshmikanth2017mass}
Lakshmikanth T, Olin A, Chen Y, Mikes J, Fredlund E, Remberger M, et~al.
\newblock Mass cytometry and topological data analysis reveal immune parameters
  associated with complications after allogeneic stem cell transplantation.
\newblock Cell reports. 2017;20(9):2238--2250.

\bibitem{rizvi2017single}
Rizvi AH, Camara PG, Kandror EK, Roberts TJ, Schieren I, Maniatis T, et~al.
\newblock Single-cell topological RNA-seq analysis reveals insights into
  cellular differentiation and development.
\newblock Nature Biotechnology. 2017;35(6):551–560.
\newblock doi:{10.1038/nbt.3854}.

\bibitem{buchet2016efficient}
Buchet M, Chazal F, Oudot SY, Sheehy DR.
\newblock Efficient and robust persistent homology for measures.
\newblock Computational Geometry. 2016;58:70--96.
\newblock doi:{https://doi.org/10.1016/j.comgeo.2016.07.001}.

\bibitem{cohen2007stability}
Cohen-Steiner D, Edelsbrunner H, Harer J.
\newblock Stability of Persistence Diagrams.
\newblock In: Proceedings of the Twenty-First Annual Symposium on Computational
  Geometry. SCG '05. New York, NY, USA: Association for Computing Machinery;
  2005. p. 263–271.
\newblock Available from: \url{https://doi.org/10.1145/1064092.1064133}.

\bibitem{carlsson2005persistence}
Carlsson G, Zomorodian A, Collins A, Guibas LJ.
\newblock Persistence barcodes for shapes.
\newblock International Journal of Shape Modeling. 2005;11(02):149--187.

\bibitem{chazal}
Chazal F, Cohen-Steiner D, Mérigot Q.
\newblock Geometric Inference for Probability Measures.
\newblock Foundations of Computational Mathematics. 2011;11(6):733–751.
\newblock doi:{10.1007/s10208-011-9098-0}.

\bibitem{buchet2017declutter}
Buchet M, Dey TK, Wang J, Wang Y.
\newblock Declutter and resample: Towards parameter free denoising.
\newblock In: 33rd International Symposium on Computational Geometry, SoCG
  2017. Schloss Dagstuhl, Leibniz-Zentrum f{\"u} Informatik GmbH; 2017. p.
  231--2316.

\bibitem{bellman1966dynamic}
Bellman R.
\newblock Dynamic Programming.
\newblock Science. 1966;153(3731):34--37.
\newblock doi:{10.1126/science.153.3731.34}.

\bibitem{kerber2017geometry}
Kerber M, Morozov D, Nigmetov A.
\newblock Geometry Helps to Compare Persistence Diagrams.
\newblock ACM J Exp Algorithmics. 2017;22.
\newblock doi:{10.1145/3064175}.

\bibitem{takemoto2006cutting}
Takemoto N, Intlekofer AM, Northrup JT, Wherry EJ, Reiner SL.
\newblock Cutting Edge: IL-12 inversely regulates T-bet and eomesodermin
  expression during pathogen-induced CD8+ T cell differentiation.
\newblock The Journal of Immunology. 2006;177(11):7515--7519.

\bibitem{scholzen2000ki}
Scholzen T, Gerdes J.
\newblock The Ki-67 protein: From the known and the unknown.
\newblock Journal of Cellular Physiology. 2000;182(3):311--322.
\newblock
  doi:{https://doi.org/10.1002/(SICI)1097-4652(200003)182:3<311::AID-JCP1>3.0.CO;2-9}.

\bibitem{knox2014characterization}
Knox JJ, Cosma GL, Betts MR, McLane LM.
\newblock Characterization of T-Bet and Eomes in Peripheral Human Immune Cells.
\newblock Frontiers in Immunology. 2014;5:217.
\newblock doi:{10.3389/fimmu.2014.00217}.

\bibitem{thibult2013pd}
Thibult ML, Mamessier E, Gertner-Dardenne J, Pastor S, Just-Landi S, Xerri L,
  et~al.
\newblock PD-1 is a novel regulator of human B-cell activation.
\newblock International immunology. 2013;25(2):129--137.

\bibitem{fimmu}
Wang Y, Hu J, Li Y, Xiao M, Wang H, Tian Q, et~al.
\newblock The Transcription Factor TCF1 Preserves the Effector Function of
  Exhausted CD8 T Cells During Chronic Viral Infection.
\newblock Frontiers in Immunology. 2019;10:169.
\newblock doi:{10.3389/fimmu.2019.00169}.

\bibitem{dey_wang_2019}
Dey TK, Wang Y.
\newblock Computational Topology for Data Analysis.
\newblock Cambridge University Press; 2022.
\newblock Available from: \url{https://books.google.com/books?id=PWtYEAAAQBAJ}.

\bibitem{van2015flowsom}
Van~Gassen S, Callebaut B, Van~Helden MJ, Lambrecht BN, Demeester P, Dhaene T,
  et~al.
\newblock FlowSOM: Using self-organizing maps for visualization and
  interpretation of cytometry data.
\newblock Cytometry Part A. 2015;87(7):636--645.

\bibitem{quintelier2021analyzing}
Quintelier K, Couckuyt A, Emmaneel A, Aerts J, Saeys Y, Van~Gassen S.
\newblock Analyzing high-dimensional cytometry data using FlowSOM.
\newblock Nature Protocols. 2021; p. 1--27.

\bibitem{chen2016xgboost}
Chen T, Guestrin C.
\newblock XGBoost: A Scalable Tree Boosting System.
\newblock In: Proceedings of the 22nd ACM SIGKDD International Conference on
  Knowledge Discovery and Data Mining. KDD '16. New York, NY, USA: Association
  for Computing Machinery; 2016. p. 785–794.
\newblock Available from: \url{https://doi.org/10.1145/2939672.2939785}.

\bibitem{gudhi:urm}
{The GUDHI Project}.
\newblock {GUDHI} User and Reference Manual.
\newblock {3.4.1} ed. {GUDHI Editorial Board}; 2021.
\newblock Available from: \url{https://gudhi.inria.fr/doc/3.4.1/}.

\bibitem{sklearn_api}
Buitinck L, Louppe G, Blondel M, Pedregosa F, Mueller A, Grisel O, et~al.
\newblock {API} design for machine learning software: experiences from the
  scikit-learn project.
\newblock In: ECML PKDD Workshop: Languages for Data Mining and Machine
  Learning; 2013. p. 108--122.

\bibitem{mit}
Panchenko D. Statistics for Applications: 18.650; 2006.
\newblock Available from: \url{https://ocw.mit.edu}.

\bibitem{bernas2008quadratic}
Bernas T, Asem EK, Robinson JP, Rajwa B.
\newblock Quadratic form: a robust metric for quantitative comparison of flow
  cytometric histograms.
\newblock Cytometry Part A: the journal of the International Society for
  Analytical Cytology. 2008;73(8):715--726.

\bibitem{Hafner}
Hafner J, Sawhney HS, Equitz W, Flickner M, Niblack W.
\newblock Efficient color histogram indexing for quadratic form distance
  functions.
\newblock IEEE Transactions on Pattern Analysis and Machine Intelligence.
  1995;17(7):729--736.
\newblock doi:{10.1109/34.391417}.

\end{thebibliography}

\newpage
\appendix
\renewcommand{\thefigure}{S\arabic{figure}}
\setcounter{figure}{0}
\renewcommand{\thesection}{S\arabic{section}}
\section{Some omitted details}
We have already introduced the concept of persistent homology. Its complete exposition is beyond the scope of this paper. We provide another intuitive example to illustrate its application in capturing prominent features hidden in a PCD. We refer to \cite{edelsbrunner2010computational,dey_wang_2019} for a detailed exposition of 
topological persistence.

\begin{figure}[!hbt]
    \centering
    \includegraphics[width=0.9\linewidth]{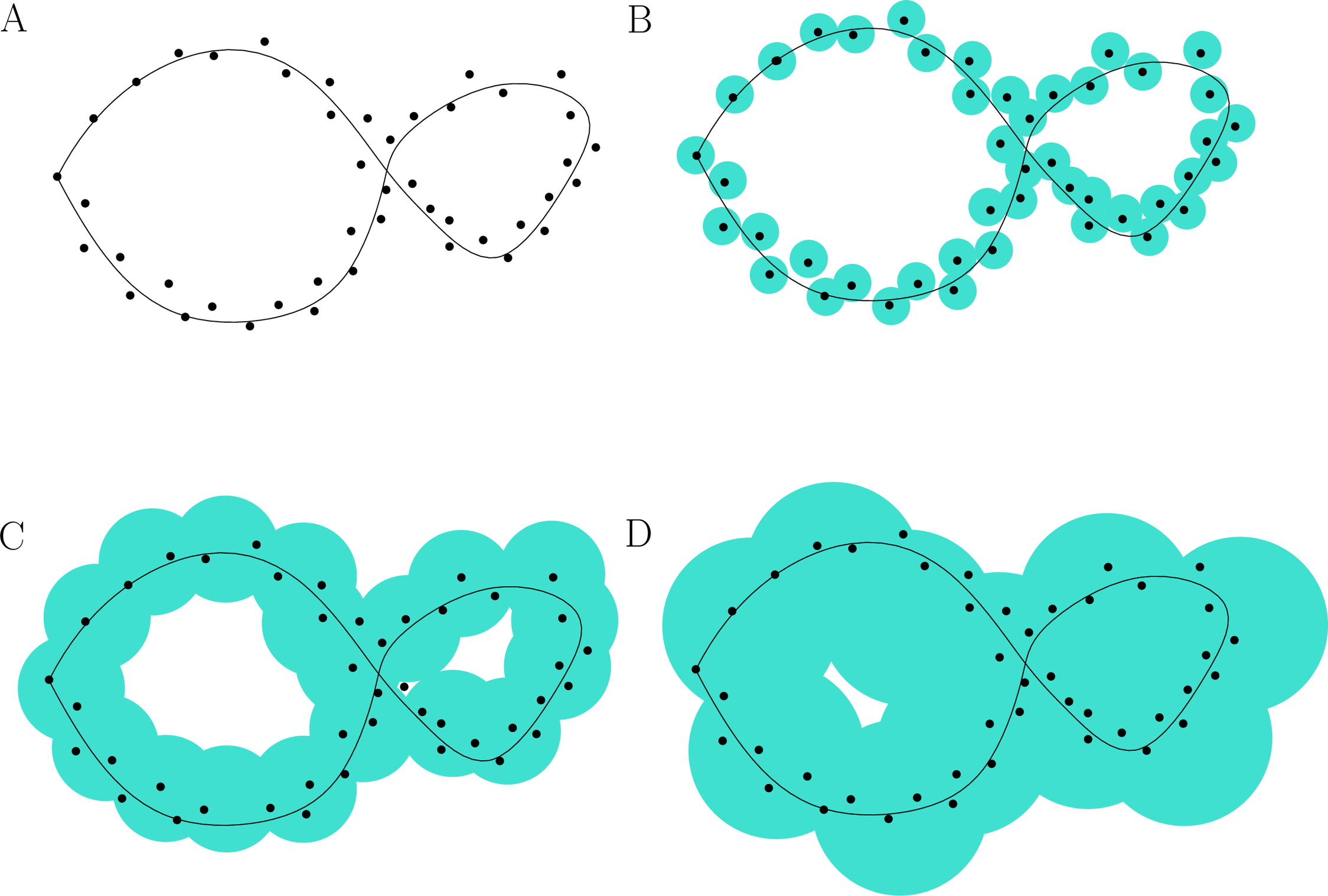}
    \caption{{\bf Intuition of persistence.} \textbf{(A)} A set of points $P$ sampled from a curve. {\bf (B)} An Euclidean ball of radius $\epsilon$ is grown around each point in $P$. {\bf (C)} As $\epsilon$ increases the smaller hole gets filled up. {\bf (D)} The larger hole still `persists' even though the smaller hole gets filled. Figures are adopted from \cite[Fig.~4.2]{dey_wang_2019}.}
    \label{fig:s1}
\end{figure}

Suppose a set of points $P$ is sampled along a curve with two `holes'; see Fig.~\ref{fig:s1}. If we grow balls of radius $\epsilon$ starting from zero around the sampled points we see that different holes get filled up at different times. The bigger hole gets filled at a much later time than the small one and the spurious ones. Persistent homology formalizes this idea of tracking the lifetime of topological features (homology groups). 
\begin{definition}[Simplex]
A $k$-simplex $\sigma$ is the convex hull of $k+1$ affinely independent point set $P$. We call $k$ to be the dimension of $\sigma$ and denote \emph{dim}$(\sigma) = k$. A \textit{face}  $\sigma^{\prime}$ of $\sigma$ is the convex hull of non-empty subset of $P$ and this relation is given by $\sigma^{\prime} \subseteq \sigma$.
\end{definition}
In particular $0$-simplex is a point, $1$-simplex is an edge, $2$-simplex is a triangle and so on.
\begin{definition}[Simplicial Complex]
A set of simplices is defined as a simplicial complex $\mathcal{K}$ if the following two restrictions hold
\begin{itemize}
    \item If $\sigma \in \mathcal{K}$ and $\sigma^{\prime} \subseteq \sigma$ then $\sigma^{\prime} \in \mathcal{K}$.
    \item For any two simplices $\sigma, \tau \in \mathcal{K}$, $\sigma \cap \tau$ is either empty or a face of both $\sigma$ and $\tau$. 
\end{itemize}
The \emph{dimension} of a simplicial complex $\mathcal{K}$ is the maximum dimension of any of its simplices.
\end{definition}
\begin{definition}[Filtration]
\label{def:fil}
Given a Simplicial Complex $\mathcal{K}$ and a \textit{monotonic} function $f:\mathcal{K} \rightarrow \mathbb{R}$ we get a nested sequence of subcomplex by denoting $\mathcal{K}_{i} = f^{-1}(-\infty, a_i]$ which is defined as \textit{filtration}: $$\phi = \mathcal{K}_0 \subseteq \mathcal{K}_1 \subseteq \mathcal{K}_2 \subseteq \ldots \subseteq \mathcal{K}_n = \mathcal{K}$$
By \textit{monotonic} we mean that if $\sigma^{\prime} \subseteq \sigma$, $f(\sigma^{\prime}) \leq f(\sigma)$.
\end{definition}

\section{Algorithms}
\label{sec:alg}
\renewcommand{\thealgorithm}{\Alph{algorithm}}
\begin{algorithm}[!t]
\centering
\caption{DisPers}\label{alg:dispers}
\renewcommand{\algorithmicrequire}{\textbf{Input:}}
\renewcommand{\algorithmicensure}{\textbf{Output:}}
\renewcommand{\algorithmicrepeat}{\textbf{begin}}
\renewcommand{\algorithmicuntil}{\textbf{end}}

\begin{algorithmic}[1]
\Require {$c$: Cytometry dataset, $k$: Nearest neighbors to consider, $n$:  Number of samples}
\Ensure{ $Dgm(c^\prime)$: Persistence diagram of $c^\prime$ sampled from cytometry data $c$}
\Repeat
\State Subsample $c$
\State $c^{\prime} \gets$ $n$ samples from $c$
\State $G \gets$ GEN-COMPLETE-GRAPH($c^{\prime}$, $k$)
\State $Dgm(c^\prime) \gets$ COMPUTE-PERSISTENCE($G$)
\State \Return{$Dgm(c^\prime)$}
\Until
\end{algorithmic}
\end{algorithm}
Since a graph is a 1-dimensional simplicial complex, for a graph with $m$ edges and $n$
vertices, we can compute its $0$-dim persistence in $\mathcal{O}(m \log n)$ time with Kruskal like minimum spanning tree algorithm \cite{edelsbrunner2010computational}.

\begin{algorithm}[!htb]
\centering
\caption{Gen-Complete-Graph}\label{alg:genvert}
\renewcommand{\algorithmicrequire}{\textbf{Input:}}
\renewcommand{\algorithmicensure}{\textbf{Output:}}
\renewcommand{\algorithmicrepeat}{\textbf{begin}}
\renewcommand{\algorithmicuntil}{\textbf{end}}

\begin{algorithmic}[1]
\Require {$c$: Sampled cytometry data, $k$: Nearest neighbors to consider}
\Ensure{ $G(V, E)$: Complete Weighted Graph on $c$}
\Procedure{GEN-COMPLETE-GRAPH}{}
\State $G(V,E) \gets \phi$ \Comment{$V$ is the set of nodes and $E$ is the set of edges of $G$}
\ForAll{$v \in {c}$}
\State Compute $v_1, v_2, \ldots, v_k $, k-nearest neighbors of $v$
\State $w(v) \gets -\frac{1}{k}\sqrt{\sum_{i}\Vert v-v_i\Vert^{2}}$ \Comment{$w(v)$ denotes vertex weight}
\State $V(G) \gets V(G) \bigcup \text{ }\big(v, w(v)\big)$
\EndFor
\ForAll{$\{u, v\} \in c \times c$ and $u \neq v$}
\State $e \gets \{u, v\}$
\State $w(e) \gets \Vert u - v \Vert$ \Comment{$w(e)$ denotes weight of the edge}
\State $E(G) \gets E(G) \bigcup \text{ }(e, w(e))$
\EndFor
\State \Return{$G(V, E)$}
\EndProcedure
\end{algorithmic}
\end{algorithm}
Algorithm~\ref{alg:perscomp} shows the steps of Persistence computation. Consider a generic step when an edge $e \in G$ is introduced in the minimum spanning forest that joins two forests rooted at nodes $v_0$ and $v_1$. For the new tree we will choose the root as node having smaller weight and the edge $e$ pairs with the node having larger weight. Essentially we are choosing to kill the youngest connected component created by, with a little abuse of notation, the vertex $argmax(w(v_0), w(v_1))$. We define persistence of the edge as 
\begin{equation}
    p(e)= w(e) - \max\{w(v_0),w(v_1)\}
\end{equation}
It is important to point out that an edge may or may not necessarily kill a connected component. If it does not kill a connected component it definitely creates a $1$-cycle and we pair the edge with a special vertex with $w(v)=\infty$ and define $pers(e) = \infty$.

\begin{algorithm}[!htb]
\centering
\caption{Compute Persistence Diagram}\label{alg:perscomp}
\renewcommand{\algorithmicrequire}{\textbf{Input:}}
\renewcommand{\algorithmicensure}{\textbf{Output:}}
\renewcommand{\algorithmicrepeat}{\textbf{begin}}
\renewcommand{\algorithmicuntil}{\textbf{end}}

\begin{algorithmic}[1]
\Require {$G(V, E)$: Complete weighted graph on sampled cytometry data $c$}
\Ensure{ $Dgm(c)$: Persistence Diagram}
\Procedure{COMPUTE-PERSISTENCE}{}
\State $E^{\prime} \gets$ Sort $E$ in increasing order of $w(e)$ with $e \in E$
\State $P_0 \gets \phi$ \Comment{$P_0$ tracks $0$-dim birth and death}
\State $P_1 \gets \phi$ \Comment{$P_1$ tracks $1$-dim birth}
\ForAll{$e = (u, v) \in {E^{\prime}}$}
\State $Root_0 \gets find(u)$
\State $Root_1 \gets find(v)$
\If{$Root_0\neq Root_1$}
\State $birth \gets max\{w(Root_0),w(Root_1)\}$ \Comment{$e$ kills youngest homology class}
\State $death \gets w(e)$
\State $pers(e) \gets death - birth$
\State $merge(Root_0,Root_1)$
\State $P_0 \gets P_0 \cup (birth, death)$
\Else
\State $P_1 \gets P_1 \cup (w(e), \infty)$ \Comment{$e$ is a creator and creates a 1-cycle}
\EndIf
\EndFor
\State $Dgm(c) \gets \{P_0, P_1\}$
\State \Return{$Dgm(c)$}
\EndProcedure
\end{algorithmic}
\end{algorithm}

\section{Supplementary Figures}
\begin{figure}[!hbt]
    \centering
    \includegraphics[width=0.9\linewidth]{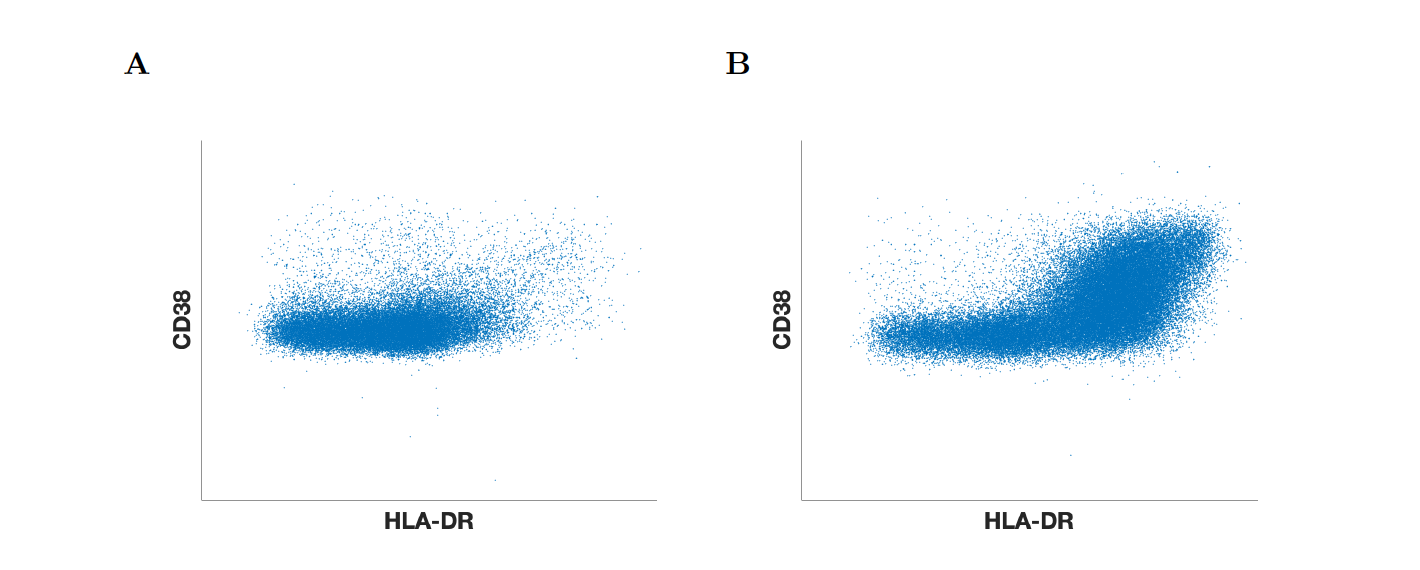}
    \caption{{\bf Scatter plots demonstrating change in structure of cytometry data.} Transformed scatter plot for HLA-DR/CD38 axes for CD8+ T cell PCD in a singular \textbf{(A)} healthy donor and \textbf{(B)} COVID-19 patient. This plot demonstrates the `elbow' found by the authors in \cite{mathew2020deep}. The x-axis is asinh(HLA-DR/200) and the y-axis is asinh(CD38/500)}
    \label{fig:s2}
\end{figure}

\begin{figure}[!htb]
    \centering
\includegraphics[width=\linewidth]{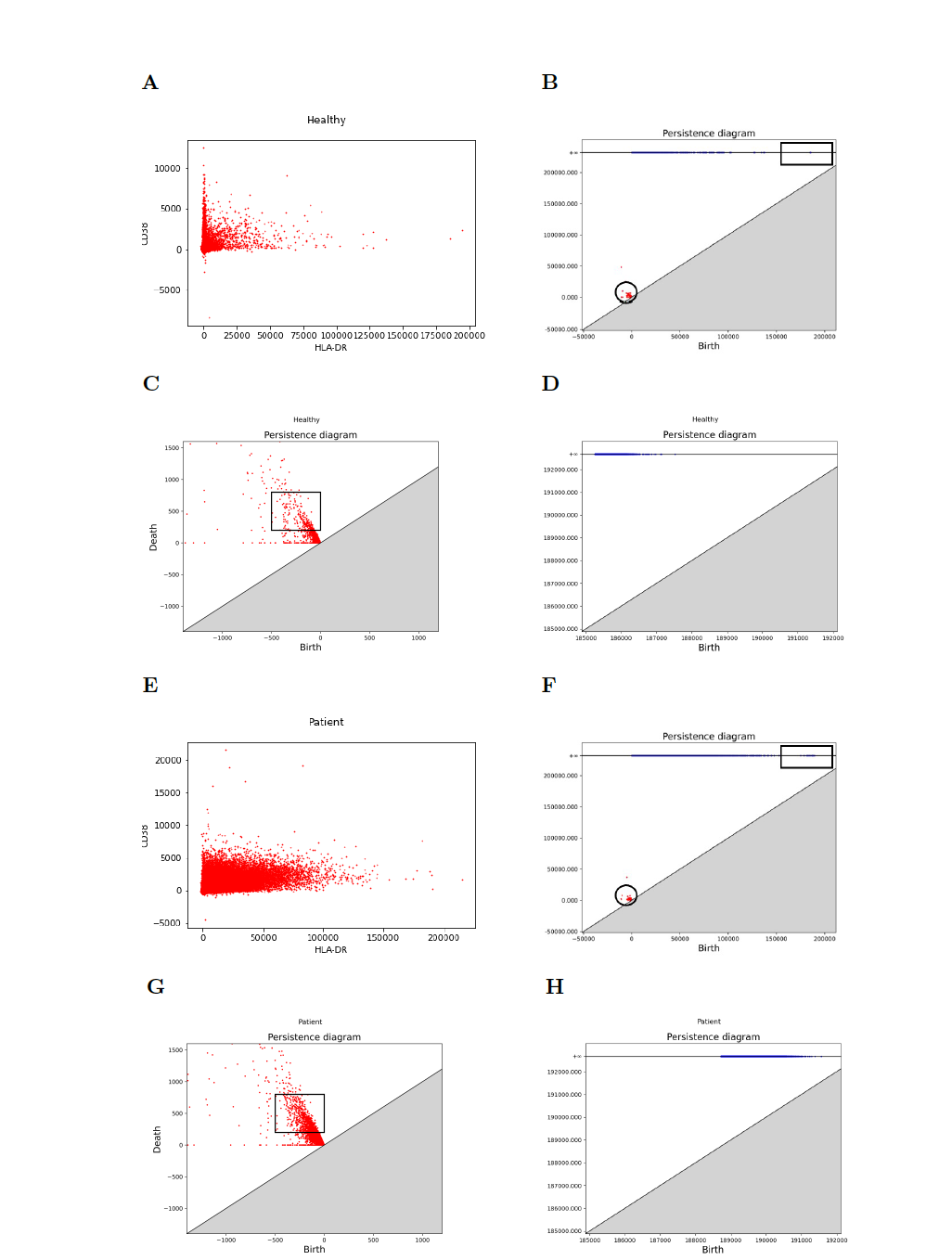}
\caption{{\bf Persistence calculations and comparisons for HLA-DR/CD38 axes for CD8+ T cell PCDs shown in Mathew et. al.\cite{mathew2020deep}} \textbf{(A)} Point cloud for individual healthy control in HLA-DR/CD38 expression levels. \textbf{(B)} Complete persistence diagram for the healthy control shown in (A). Boxes indicate zoomed regions for figures (C) and (D); \textbf{(C)} Zoomed in region from (B) of ${\sf H}_0$ persistence diagram. Box shows area of low density compared to patient persistence diagram; \textbf{(D)} Zoomed in region from (B) of ${\sf H}_1$ persistence diagram; \textbf{(E)-(H)} Same as (A)-(D), but for an individual COVID-19 patient. The box in (G) is more densely populated than the identical box in (C).}
\label{fig:s3}
\end{figure}

\begin{figure}[!htb]
    \centering
    \includegraphics[width=\linewidth]{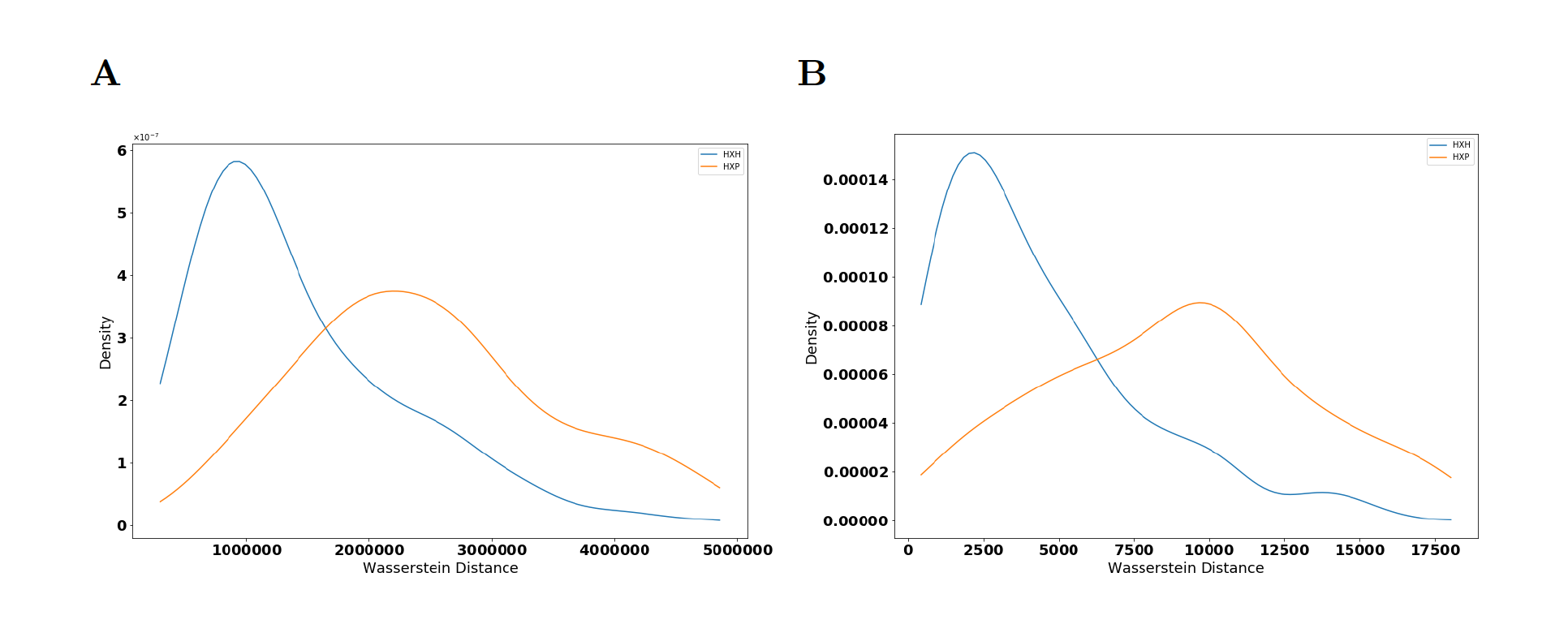}
    \caption{{\bf Distributions of Wasserstein distances between persistence diagrams calculated from 200 pairs of individuals}. Distributions of Wasserstein distances between \textbf{(A)} $\sf{H}_0$-persistence diagrams ($p=6.75\times 10^{-20}$, QFD=$0.190$) and \textbf{(B)} $\sf{H}_1$-persistence diagrams ($p=4.74\times 10^{-24}$, QFD=$0.220$) for CD8+ T cells. Distances between pairs of healthy controls (H $\times$ H) and pairs of a healthy control and a COVID-19 patient (H $\times$ P) are overlaid. Persistence diagrams are calculated from point clouds in the T-bet, Eomes, and Ki-67 axes. This figure plots distributions of \textbf{200} randomly selected pairs, while Fig.~\ref{fig:wasserstein} plots distributions of 100 randomly selected pairs.}
    \label{fig:s4}
\end{figure}

\begin{figure}[!htb]
    \centering
    \includegraphics[width=\linewidth]{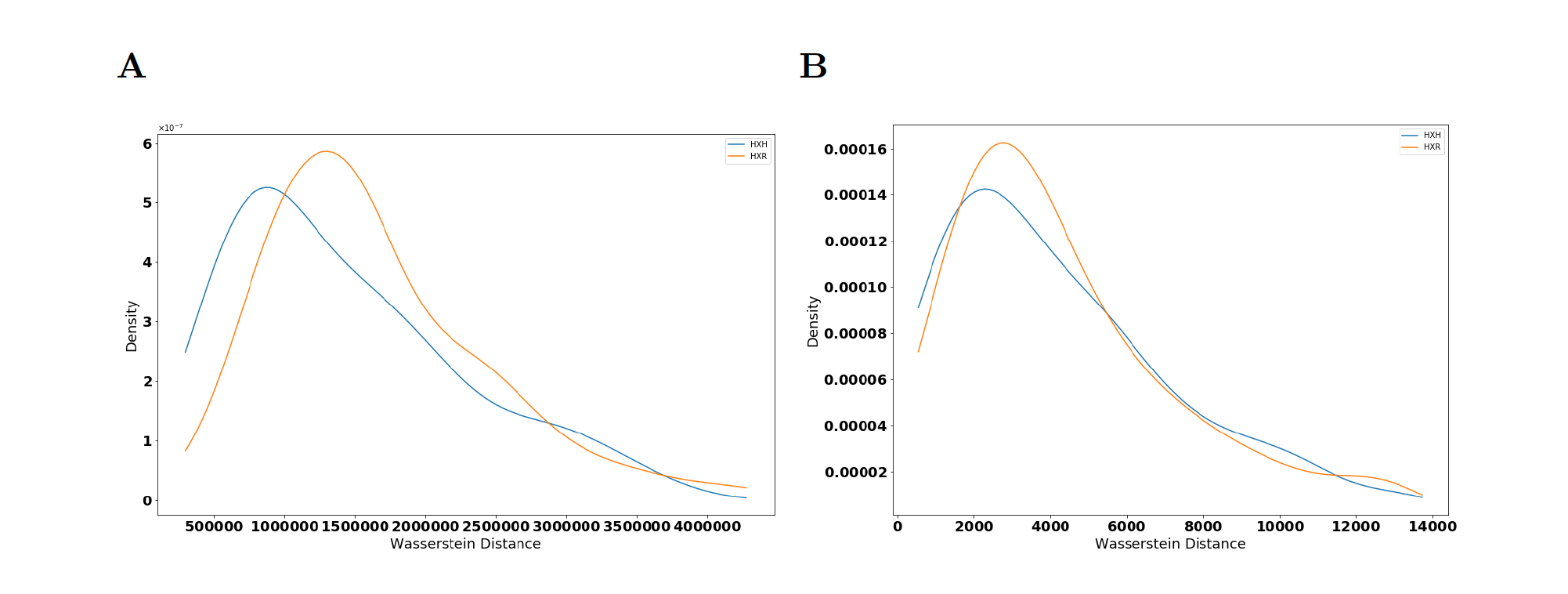}
    \caption{{\bf Distributions of Wasserstein distances between persistence diagrams for healthy controls and recovered individuals calculated using 3 most important proteins for XGBoost classification of CD8+ T cells from healthy or infected individuals}. Distributions of Wasserstein distances between \textbf{(A)} $\sf{H}_0$-persistence diagrams ($p=0.131$, QFD=$0.005$) and \textbf{(B)} $\sf{H}_1$-persistence diagrams ($p=0.344$, QFD=$0.001$) for CD8+ T cells. Distances between pairs of healthy controls (H $\times$ H) and pairs of a healthy control and a individual that recovered from COVID-19 (H $\times$ R) are overlaid. Persistence diagrams are calculated from point clouds in the T-bet, Eomes and Ki-67 axes. p-values are calculated from a 2-sided KS test.}
    \label{fig:s5}
\end{figure}

\begin{figure}[!htb]
    \centering
    \includegraphics[width=\linewidth]{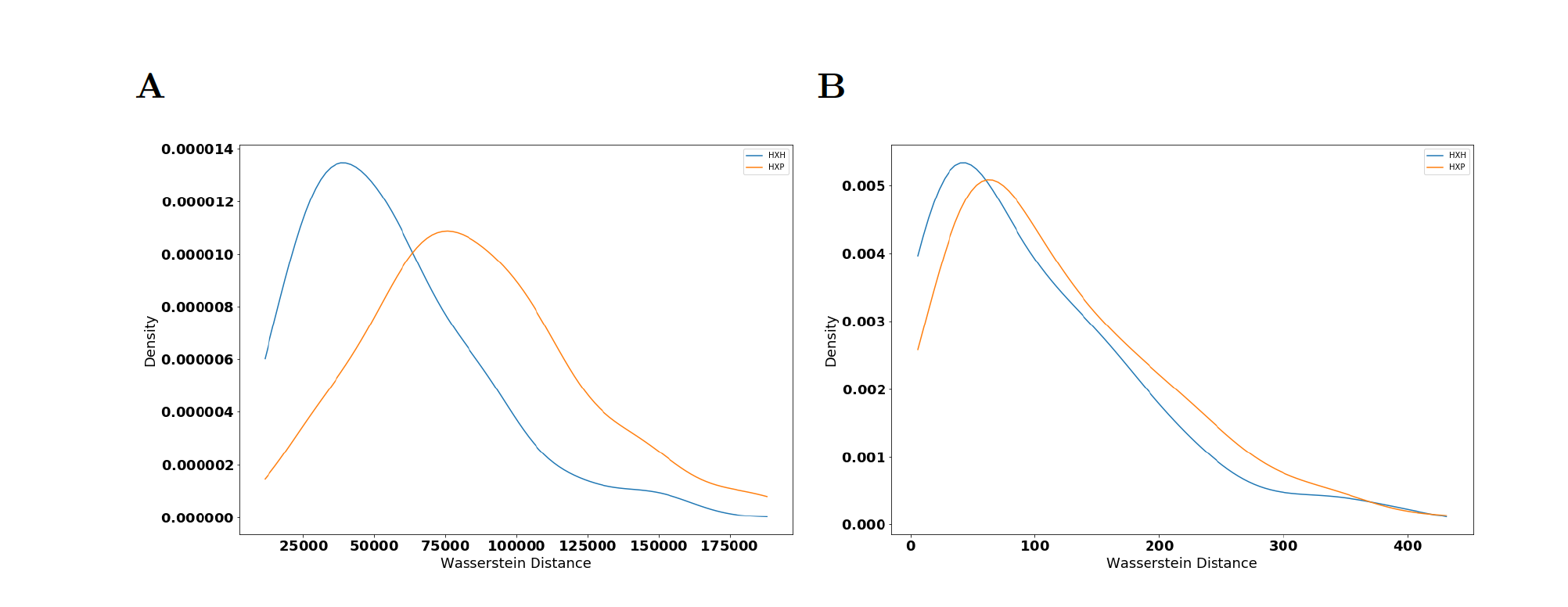}
    \caption{{\bf Distributions of Wasserstein distances between persistence diagrams calculated using 3 least important proteins for XGBoost classification of CD8+ T cells}. Distributions of Wasserstein distances between \textbf{(A)} $\sf{H}_0$-persistence diagrams ($p=2.75\times10^{-8}$, QFD=$0.051$) and \textbf{(B)} $\sf{H}_1$-persistence diagrams ($p=0.111$, QFD=$0.022$) for CD8+ T cells. Distances between pairs of healthy controls (H $\times$ H) and pairs of a healthy control and a COVID-19 patient (H $\times$ P) are overlaid. Persistence diagrams are calculated from point clouds in the IgD, CD4 and CD20 axes. p-values are calculated from a 2-sided KS test.}
    \label{fig:s6}
\end{figure}

\begin{figure}[!htb]
    \centering
    \includegraphics[width=\linewidth]{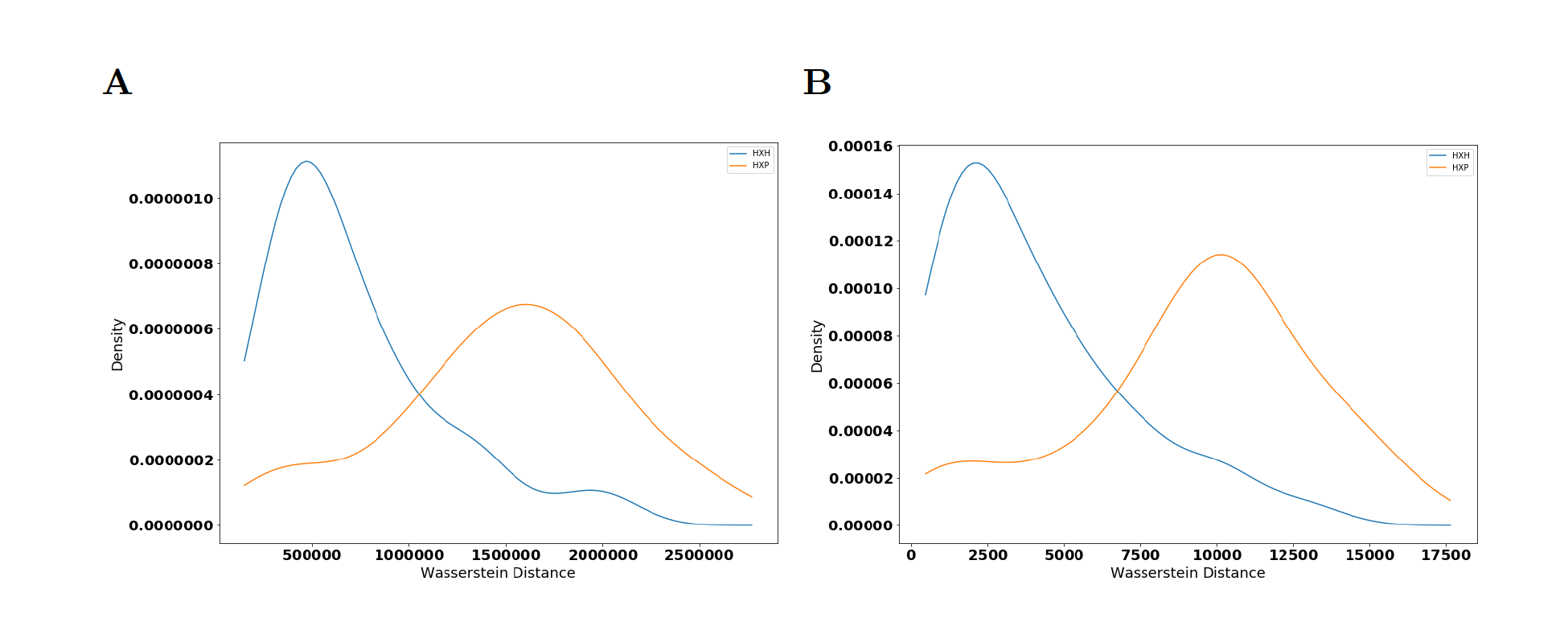}
    \caption{{\bf Distributions of Wasserstein distances between persistence diagrams calculated using 2 most important proteins for XGBoost classification of CD8+ T cells from healthy or infected individuals}. Distributions of Wasserstein distances between \textbf{(A)} $\sf{H}_0$-persistence diagrams ($p=6.31\times10^{-19}$, QFD=$0.261$) and \textbf{(B)} $\sf{H}_1$-persistence diagrams ($p=6.31\times10^{-19}$, QFD=$0.276$) for CD8+ T cells. Distances between pairs of healthy controls (H $\times$ H) and pairs of a healthy control and a COVID-19 patient (H $\times$ P) are overlaid. Persistence diagrams are calculated from point clouds in the T-bet and Eomes axes. p-values are calculated from a 2-sided KS test.}
    \label{fig:s7}
\end{figure}

\begin{figure}[!htb]
    \centering
    \includegraphics[width=\linewidth]{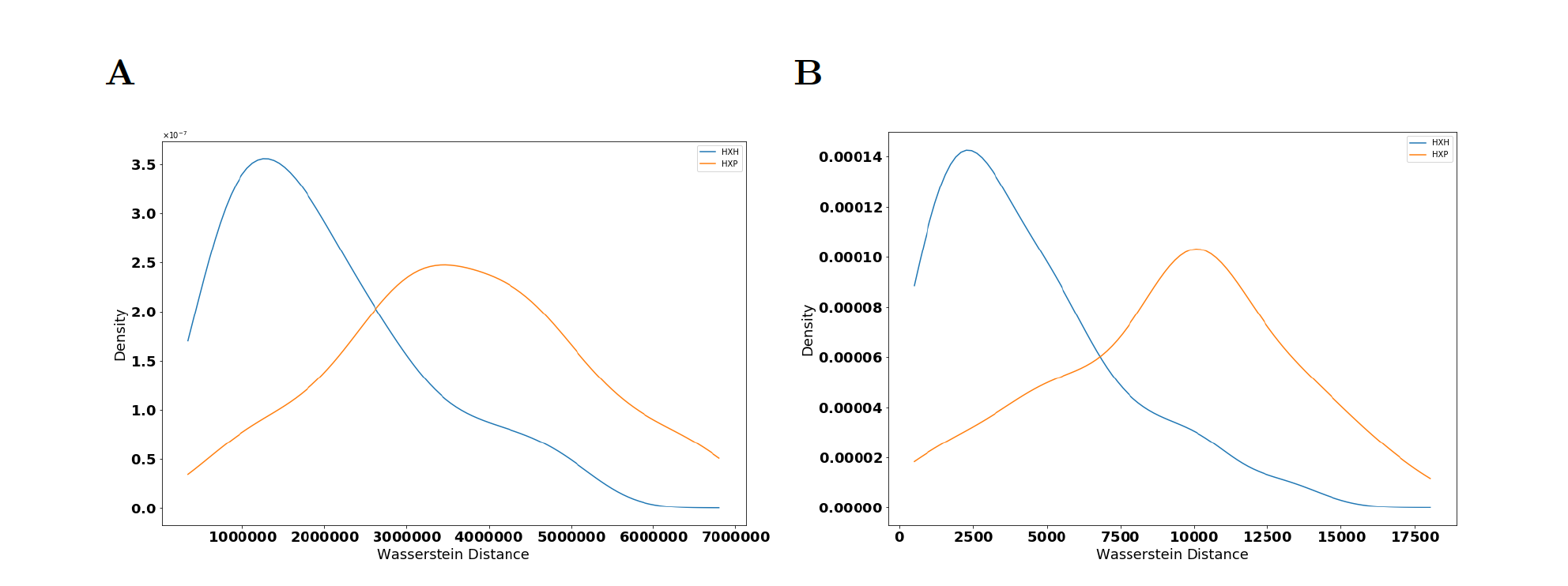}
    \caption{{\bf Distributions of Wasserstein distances between persistence diagrams calculated using 4 most important proteins for XGBoost classification of CD8+ T cells from healthy or infected individuals}. Distributions of Wasserstein distances between \textbf{(A)} $\sf{H}_0$-persistence diagrams ($p=3.35\times10^{-13}$, QFD=$0.267$) and \textbf{(B)} $\sf{H}_1$-persistence diagrams ($p=3.04\times10^{-14}$, QFD=$0.265$) for CD8+ T cells. Distances between pairs of healthy controls (H $\times$ H) and pairs of a healthy control and a COVID-19 patient (H $\times$ P) are overlaid. Persistence diagrams are calculated from point clouds in the T-bet, Eomes, Tox and TCF-1 axes. p-values are calculated from a 2-sided KS test.}
    \label{fig:s8}
\end{figure}

\begin{figure}[!htb]
    \centering
    \includegraphics[width=\linewidth]{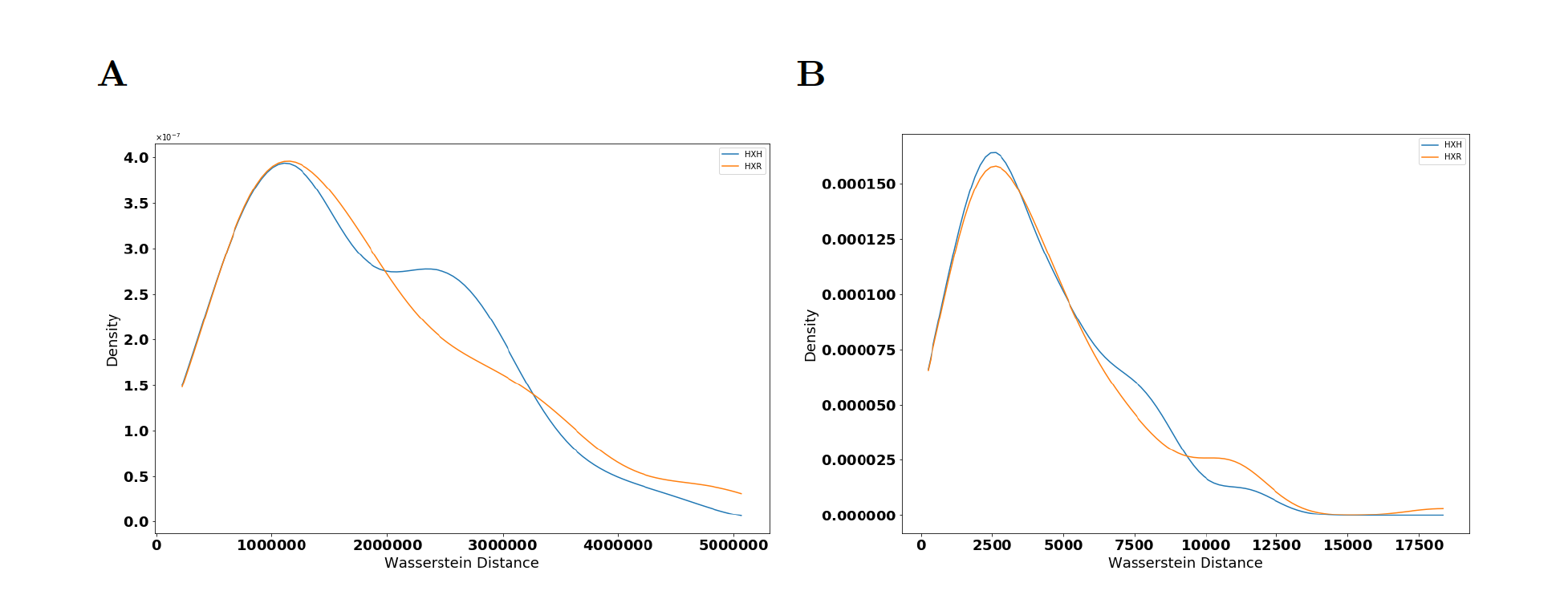}
    \caption{{\bf Distributions of Wasserstein distances between persistence diagrams calculated using 3 most important proteins for XGBoost classification of recovered CD8+ T cells}. Distributions of Wasserstein distances between \textbf{(A)} $\sf{H}_0$-persistence diagrams ($p=0.908$, QFD=$0.002$) and \textbf{(B)} $\sf{H}_1$-persistence diagrams ($p=0.994$, QFD=$0.001$) for CD8+ T cell. Distances between pairs of healthy controls (H $\times$ H) and pairs of a healthy control and a individual that recovered from COVID-19 (H $\times$ R) are overlaid. Persistence diagrams are calculated from point clouds in the CD45RA, Eomes and TCF-1 axes. These 3 proteins are the best distinguishing features for the XGBoost classifier to distinguish cells from healthy controls from those from recovered individuals. p-values are calculated from a 2-sided KS test.}
    \label{fig:s9}
\end{figure}

\begin{figure}[!htb]
    \centering
    \includegraphics[width=\linewidth]{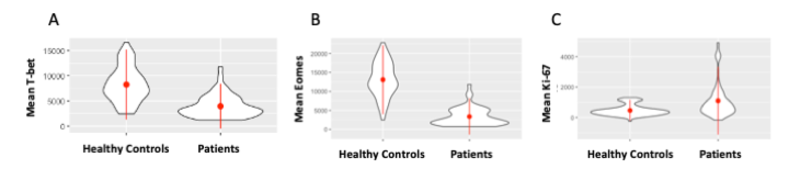}
    \caption{\textbf{Probability distribution function (pdf) of mean abundances of proteins identified by XGBoost to be important to patient classification in CD8+ T cells.} PDFs for \textbf{(A)} T-bet, \textbf{(B)} Eomes, and \textbf{(C)} Ki-67 for CD8+ T cell PCDs in each healthy control and COVID-19 patient shown using violin plots. The thickness of the ``violin" denotes the value of the pdf. Data distributions are calculated from the mean protein abundances across all non-naïve CD8+ T cells for each individual. Red dots represent the mean of the data, and red lines represent the standard deviation.}
    \label{fig:s10}
\end{figure}

\begin{figure}[!htb]
    \centering
    \includegraphics[width=\linewidth]{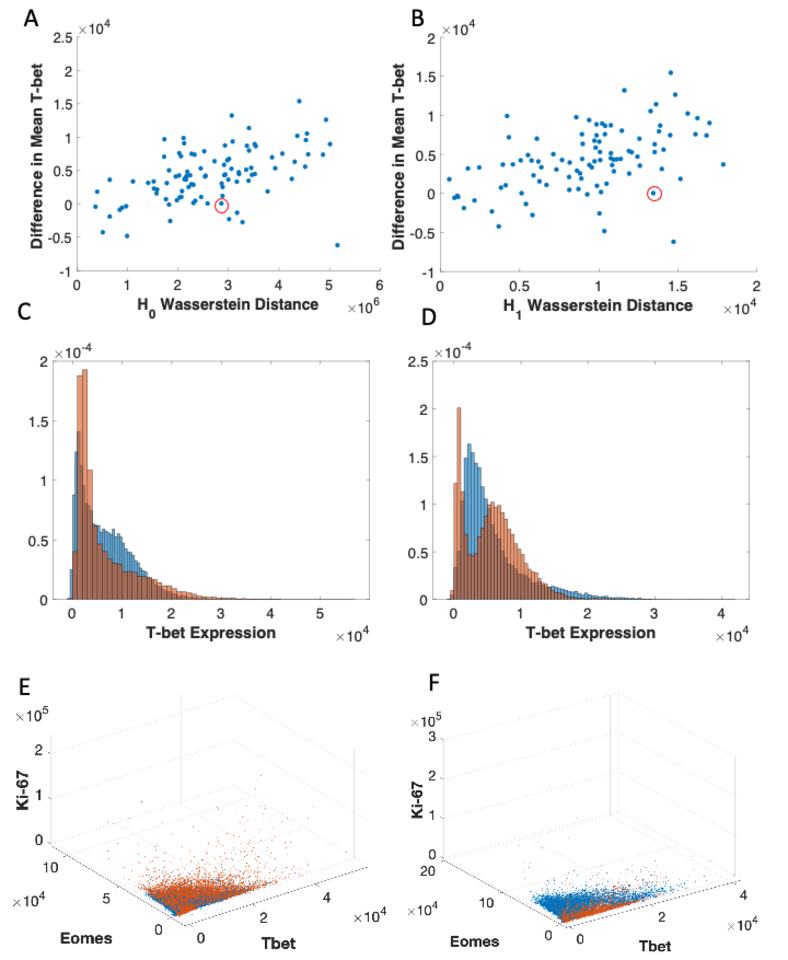}
    \caption{\textbf{Demonstration of persistent homology’s ability to capture more information than change in magnitude of single protein measurements in CD8+ T cells. (A-B)} Scatter plot showing the relationship between Wasserstein distance between $\sf{H}_0$ and $\sf{H}_1$ persistence diagrams and the difference in mean T-bet abundance for CD8+ T cells for random pairs of healthy controls and COVID-19 patients. Red circles highlight a pair of individuals that generated a large Wasserstein distance despite a small difference in mean T-bet expressions, which are further analyzed in (C-F). \textbf{(C-D)} Histograms showing the T-bet expression of non-naïve CD8+ T cells for the healthy control (blue) and COVID-19 patient (red) from the points circled above in (A-B). \textbf{(E-F)} Scatter plots showing the T-bet, Eomes, and Ki-67 abundances for the healthy control (blue) and COVID-19 patient (red) from the points circled in (A-B).}
    \label{fig:s11}
\end{figure}

\begin{figure}[!htb]
    \centering
    \includegraphics[width=\linewidth]{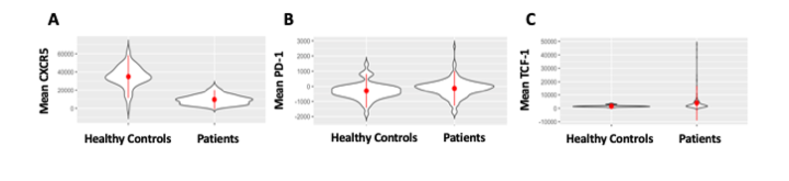}
    \caption{\textbf{Probability distribution function (pdf) of mean abundances of proteins identified by XGBoost to be important to patient classification in B cells.} PDFs for \textbf{(A)} CXCR5, \textbf{(B)} PD-1, and \textbf{(C)} TCF-1 in B cells of each healthy control and COVID-19 patient shown using violin plots. The thickness of the ``violin" denotes the value of the pdf. Data distributions are calculated from the mean protein abundances across all B cells for each individual. Red dots represent the mean of the data, and red lines represent the standard deviation.}
    \label{fig:s12}
\end{figure}

\begin{figure}[!htb]
    \centering
    \includegraphics[width=\linewidth]{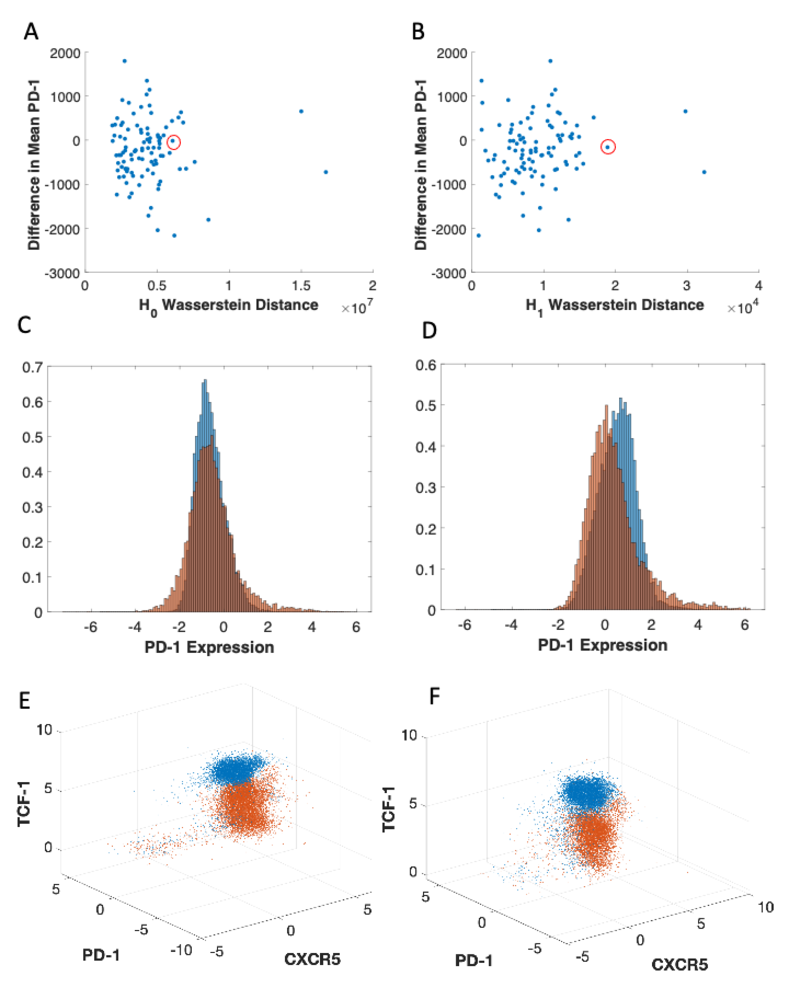}
    \caption{\textbf{Demonstration of persistent homology’s ability to capture more information than change in magnitude of single protein measurements in B cells. (A-B)} Scatter plot showing the relationship between Wasserstein distance between $\sf{H}_0$ and $\sf{H}_1$ persistence diagrams and the difference in mean PD-1 abundance for B cells for random pairs of healthy controls and COVID-19 patients. Red circles highlight a pair of individuals that generated a large Wasserstein distance despite a small difference in mean PD-1 expressions, which are further analyzed in (C-F). \textbf{(C-D)} Histograms showing the PD-1 expression of non-naïve B cells for the healthy control (blue) and COVID-19 patient (red) from the points circled above in (A-B). \textbf{(E-F)} Scatter plots showing the CXCR5, PD-1, and TCF-1 abundances for the healthy control (blue) and COVID-19 patient (red) from the points circled in (A-B). Protein expression axes in (C)-(F) are scaled to asinh(x/150), where x is the expression of the given protein.}
    \label{fig:s13}
\end{figure}

\begin{figure}[!hbt]
    \centering
    \includegraphics[width=\linewidth]{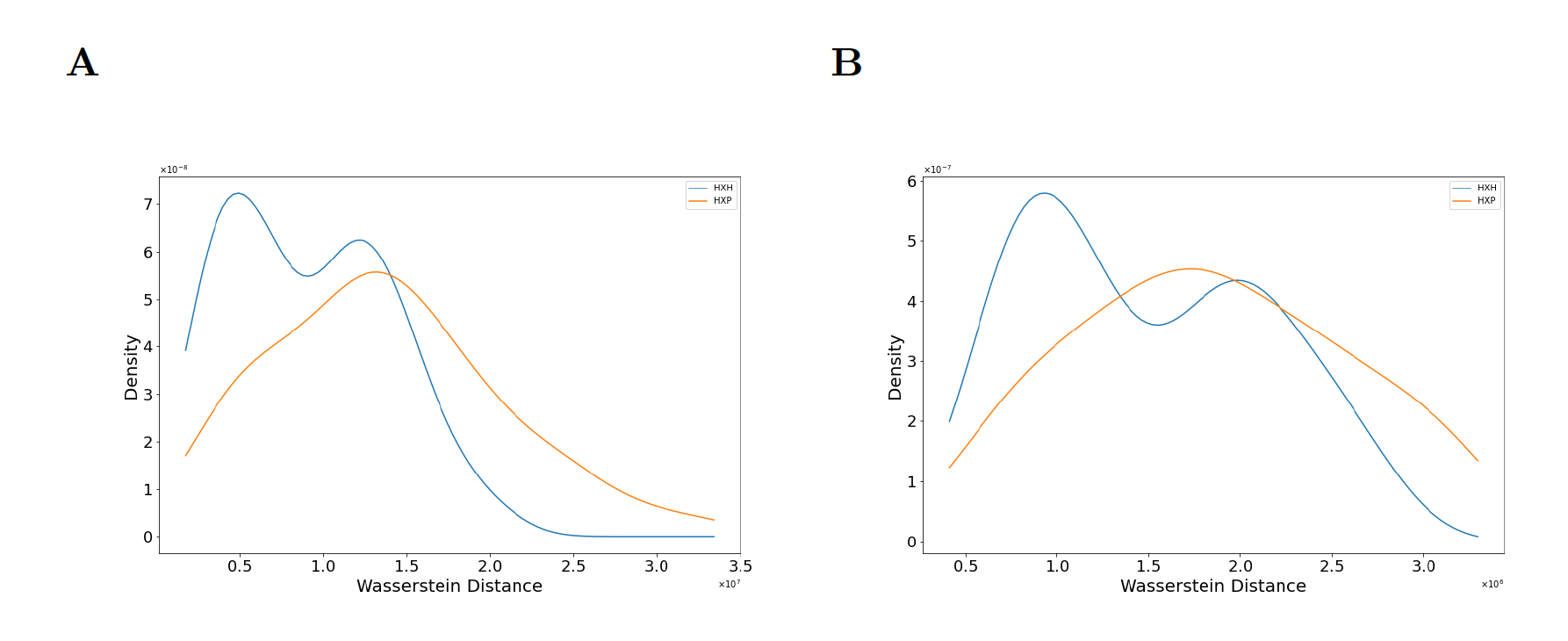}
    \caption{{\bf Distributions of Wasserstein distances between persistence diagrams calculated using Rips filtration}. {\bf(A)} Shows distributions of Wasserstein distance between ${\sf H}_0$-persistence diagrams for H$\times$H (blue line) and H$\times$P (orange line) pairs (p=$1.20 \times 10^{-4}$, QFD=$0.0575$) for CD8+ T cells. {\bf (B)} Shows distributions of Wasserstein distance between ${\sf H}_1$-persistence diagrams for H$\times$H (blue line) and H$\times$P (orange line) for the same pairs in (A) (p=$3.73 \times 10^{-3}$, QFD=$0.0343$).}
    \label{fig:s14}
\end{figure}

\begin{figure}[!htb]
    \centering
    \includegraphics[width=\linewidth]{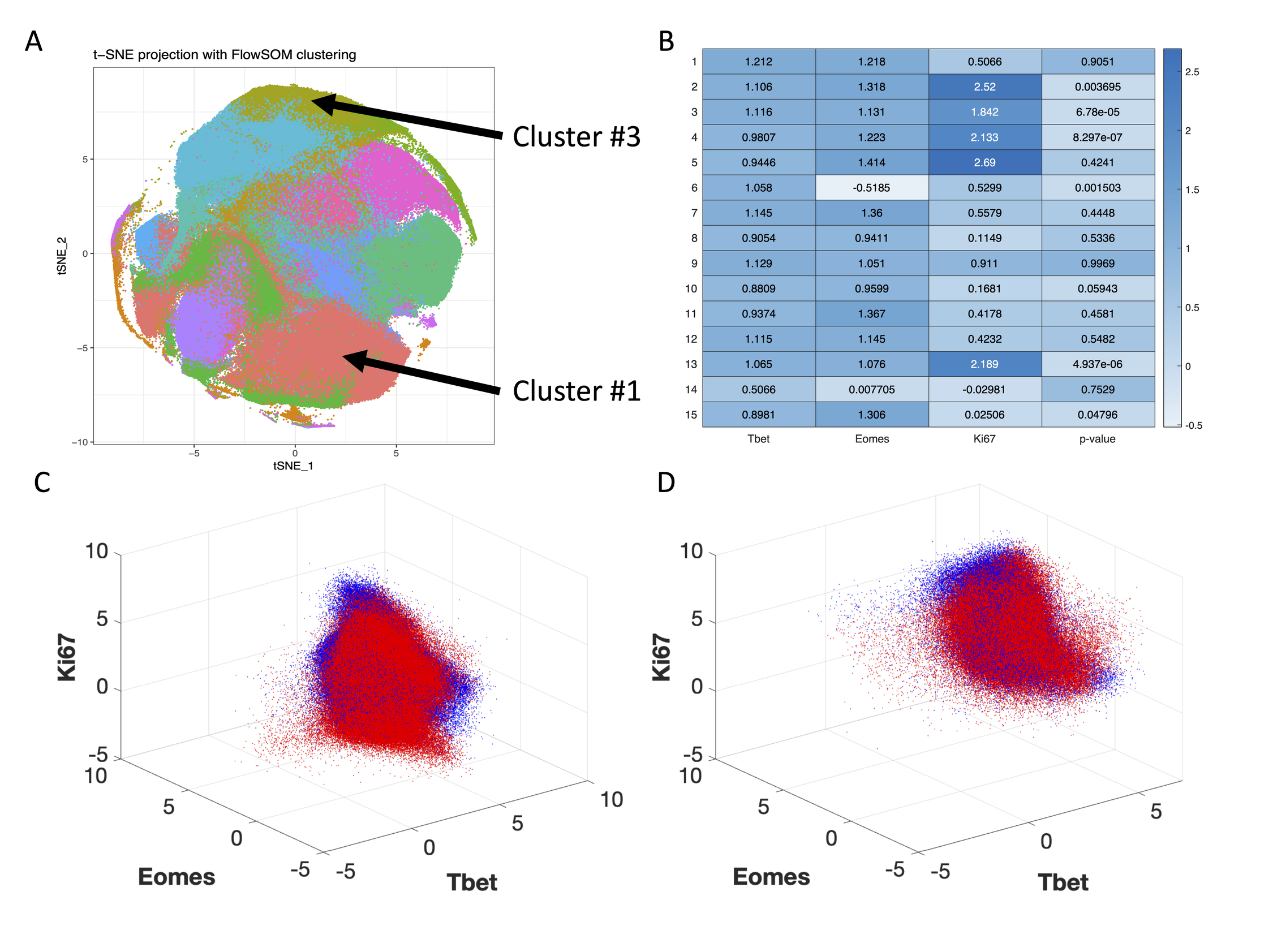}
    \caption{\textbf{Results of FlowSOM analysis for CD8+ T cells. (A)} t-SNE projection of protein expression data for CD8+ T cell PCD in Mathew et al. Each point represents a cell with 25 protein expressions. Colors represent 15 clusters identified by FlowSOM. Cluster \#1 and Cluster \#3 are selected for further topological analysis. \textbf{(B)} Heatmap showing scaled MFI for T-bet, Eomes, and Ki-67 for each cluster. Each entry in the first three columns is the MFI scaled by the average MFI of the column. The fourth column shows p-values determining differential expression of the cluster between healthy controls and COVID-19 patients. Note that Cluster \#1 has $p>0.05$ and Cluster \#3 has $p<0.05$. \textbf{(C-D)} Scatter plots showing the T-bet, Eomes, and Ki-67 abundances for all healthy controls (blue) and COVID-19 patients (red) from the cells in Cluster \#1 (C) and Cluster \#3 (D). Black circles in (C) indicate regions of single cell protein expressions which contribute to the differences in the PCD structure for the FlowSOM clusters. Axes are scaled to asinh(x/150), where x is the expression of the given protein.}
    \label{fig:s15}
\end{figure}

\begin{figure}[!htb]
    \centering
    \includegraphics[width=\linewidth]{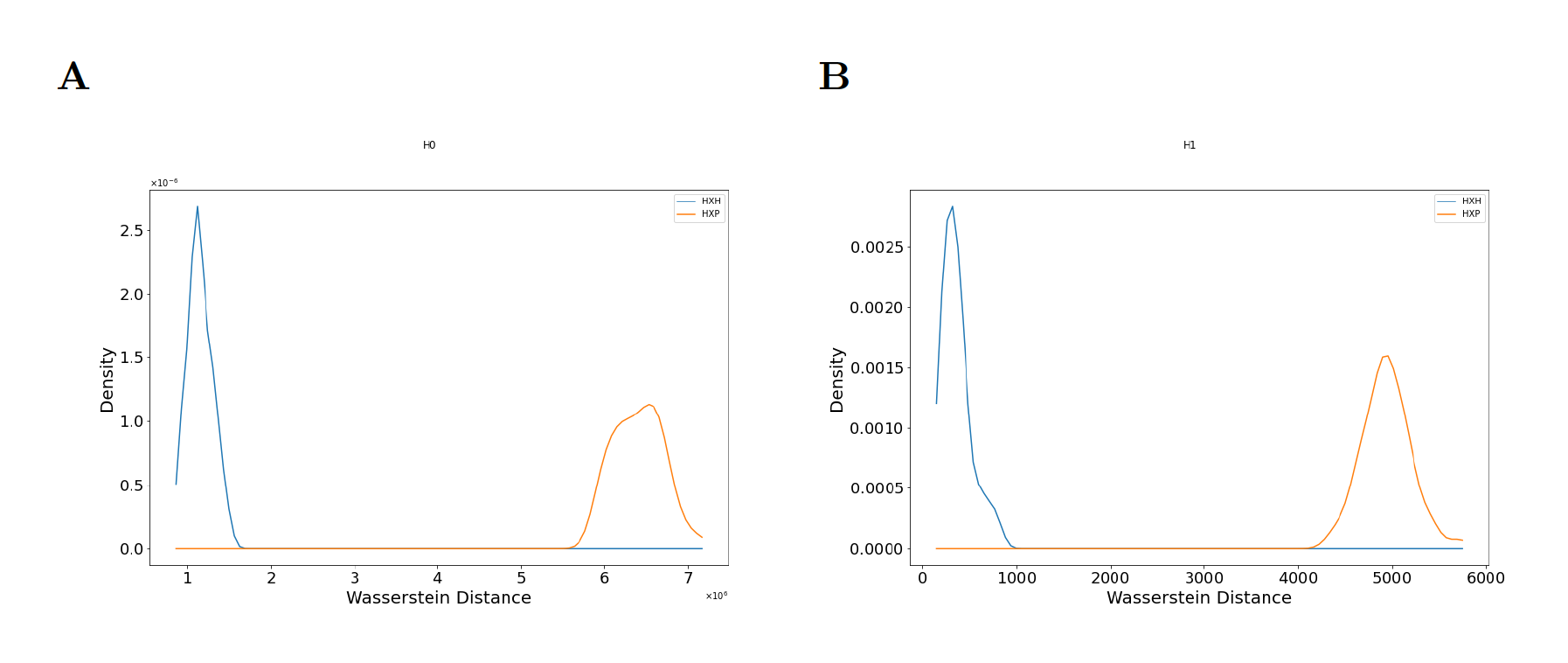}
    \caption{{\bf Distributions of Wasserstein distances between persistence diagrams from a FlowSOM cluster (Cluster {\#}3) that is differentially expressed between healthy controls and COVID-19 patients}. {\bf(A)} Shows distributions of Wasserstein distance between ${\sf H}_0$-persistence diagrams for H$\times$H (blue line) and H$\times$P (orange line) pairs (p=$2.20 \times 10^{-59}$, QFD=$1.604$). {\bf (B)} Shows distributions of Wasserstein distance between ${\sf H}_1$-persistence diagrams for H$\times$H (blue line) and H$\times$P (orange line) for the same pairs in (A) (p=$2.21 \times 10^{-59}$, QFD=$1.573$).}
    \label{fig:s16}
\end{figure}

\begin{figure}[!htb]
    \centering
    \includegraphics[width=0.95\linewidth]{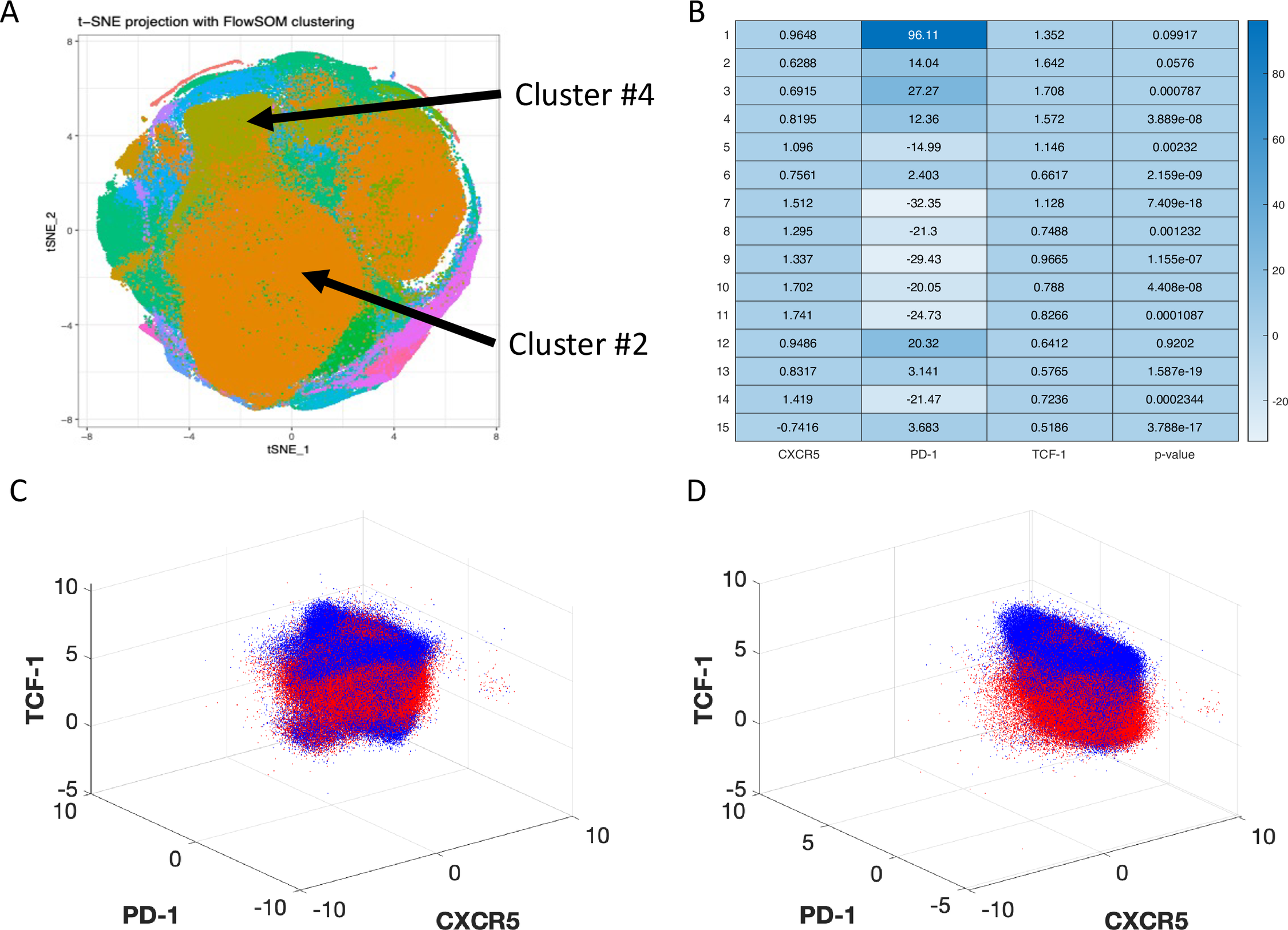}
    \caption{\textbf{Results of FlowSOM analysis for B cells. (A)} t-SNE projection of protein expression data for B cell PCD in Mathew et al. Each point represents a cell with 25 protein expressions. Colors represent 15 clusters identified by FlowSOM. Cluster \#2 and Cluster \#4 are selected for further topological analysis. \textbf{(B)} Heatmap showing scaled MFI for CXCR5, PD-1, and TCF-1 for each cluster. Each entry in the first three columns is the MFI scaled by the average MFI of the column. The fourth column shows p-values determining differential expression of the cluster between healthy controls and COVID-19 patients. Note that Cluster \#2 has $p>0.05$ and Cluster \#4 has $p<0.05$. \textbf{(C-D)} Scatter plots showing the CXCR5, PD-1, and TCF-1 abundances for all healthy controls (blue) and COVID-19 patients (red) from the cells in Cluster \#2 (C) and Cluster \#4 (D). Axes are scaled to asinh(x/150), where x is the expression of the given protein.}
    \label{fig:s17}
\end{figure}

\end{document}